\documentclass[twocolumn]{aa}
\usepackage{graphicx}
\usepackage{natbib}
\usepackage{txfonts}
\begin{document}
   \title{Extracting clean supernova spectra\thanks{Partly based on 
observations collected at the European Southern Observatory, Chile (ESO 
Programme 170.A-0519).}}	

\subtitle{Towards a quantitative analysis of high-redshift Type Ia supernova 
spectra}

   \author{S. Blondin\inst{1}
          \and
          J. R. Walsh\inst{2}
	  \and
	  B. Leibundgut\inst{1}
	  \and
	  G. Sainton\inst{3}
          }

   \offprints{S. Blondin}

   \institute{European Southern Observatory (ESO), Karl-Schwarzschild-Strasse 
2, D-85748 Garching bei M\"unchen, Germany \\
              \email{sblondin@eso.org,bleibund@eso.org}
         \and
             Space Telescope-European Coordinating Facility (ST-ECF), European
 Southern Observatory (ESO), Karl-Schwarzschild-Strasse 2, D-85748 Garching bei
 M\"unchen, Germany \\
             \email{jwalsh@eso.org}
	 \and
	     LPNHE CNRS/IN2P3, Universit\'e Paris VI \& VII, 4 Place Jussieu, 
75252 Paris Cedex 05, France \\
	     \email{sainton@in2p3.fr}
             }

   \date{Received / Accepted}

   \abstract{
We use a new technique to extract the spectrum of a supernova
from that of the contaminating background of its host galaxy, and apply it to 
the specific case of high-redshift Type Ia supernova (SN Ia) spectroscopy. The
 algorithm is based on a two-channel iterative technique employing the 
Richardson-Lucy restoration method and is implemented in the IRAF code 
\emph{specinholucy}. We run the code both on simulated (SN Ia at $z=0.5$ 
embedded in a bright host galaxy) and observed (SNe Ia at various phases up 
to $z=0.236$) data taken with VLT+FORS1 and show the advantages of using such 
a deconvolution technique in comparison with less elaborate methods. This 
paper is motivated by the need for optimal supernova spectroscopic 
 data reduction in order to make meaningful comparisons between the low and 
 high-redshift SN Ia samples. This may reveal subtle evolutionary and 
 systematic effects that could depend on redshift and bias the cosmological 
results derived from comparisons of local and high-$z$ SNe Ia in recent years.
 We describe the various aspects of the extraction in some detail as 
guidelines for the first-time user and present an optimal observing strategy 
for successful implementation of this method in future high-$z$ SN Ia 
spectroscopic follow-up programmes.
   
   }

   \maketitle
%

\section{Introduction}

The initial claim made by two independent teams -- the High-Z Supernova Search
 Team \citep{schmidt98} and the Supernova Cosmology Project \citep{P99} -- 
that the apparent dimming of Type Ia supernovae (SN Ia) at redshifts of 
$z \approx 0.5$ implies a present accelerating expansion of the universe 
\citep{R98,P99}, and its subsequent confirmation
with improved precision \citep{tonry03,knop03,barris04}, have
prompted an increased interest in these astrophysical events. On the 
theoretical side, a lot of effort and computing time has been devoted to 
multi-dimensional modelling of the explosion in order to provide physical 
input parameters for spectral synthesis calculations (see \citealt{wolfi00} 
for a review). On the observational side several teams are currently using 
SNe Ia both to probe the decelerating expansion at higher ($z \gtrsim 1$) 
redshifts -- the Hubble Higher-$z$ Supernova Search \citep{riesshz1} -- as 
well as measure second order effects (the equation-of-state parameter 
$\omega=p/\rho c^2$) at intermediate ($0.2 \lesssim z \lesssim 0.8$) redshifts
 -- such as ESSENCE \citep{miknaitis05,matheson05} and the CFHT SuperNova 
Legacy Survey, or SNLS \citep{pain02}.

However, the physical parameters governing the SN Ia explosion mechanism are 
today not fully understood and the empirical parametrisation of the entire 
SN Ia class -- the so-called ``Phillips relation'' \citep[see][]{phillips93} 
-- has no physical basis. The aforementioned observational programmes will 
detect several hundreds of high-$z$ SNe Ia, at which stage the statistical 
errors in the sample will have reached the systematic error floor. The 
cosmological effect detected via observations of Type Ia supernovae is indeed 
a subtle one (see \citealt{bruno01} for a review) and the field is limited by 
the systematic errors involved in such measurements and their possible 
evolution with redshift.

Spectroscopy is an ideal way to probe these potential evolutionary effects 
through systematic comparison of high-$z$ SN Ia spectra with local templates.
 This requires the extraction of clean SN Ia spectra devoid of background 
contamination, mainly by the host galaxy and sky lines (for ground-based 
observations). For high signal-to-noise (hereafter S/N) cases, when the 
supernova is bright with respect to its immediate underlying background, 
standard extraction software will in general work well. However, when the 
supernova is faint compared to the background, e.g. at late phases, heavily 
superposed on its host galaxy or simply at a high redshift, a more elaborate 
technique is required to ensure that we are not extracting non-SN flux.

The purpose of this paper is to test such a method that has been implemented 
in the IRAF\footnote{IRAF is distributed by the National Optical Astronomy 
Observatories, operated by the Association of Universities for Research in 
Astronomy, Inc., under contract to the National Science Foundation of the 
United States.} code \emph{specinholucy} \citep{LW03}. It is based on a 
two-channel restoration algorithm that restores a point spread function 
(PSF)-like component in a 2D image and an underlying extended background 
\emph{separately}. It is of wide astronomical use and has already been 
successfully applied to the restoration of point-source spectra in highly 
inhomogeneous backgrounds \citep[see][]{LW03}, such as is often the case for 
high-$z$ SNe Ia embedded in their host galaxy. The fact that the entire 2D 
spectrum of the background (with or without the inclusion of the restored 
point source spectrum) is restored in such an algorithm means that we have 
a firm hold on potential systematic errors associated with the restoration. 
This is done by simply comparing the residuals in the restored 2D spectrum 
with the statistical noise limit of the input 2D spectrum (the square root 
of the number of photoelectrons). Should these residuals fall below this limit
 then we can consider the restoration to be successful. This is a clear 
advantage over traditional spectral extraction routines where no secure hold 
on systematic errors is possible. Furthermore, the very nature of the 
algorithm presented in this paper implies that all non PSF-like components in 
the input 2D spectrum are automatically allocated to the background channel. 
We are making no \emph{a priori} assumptions on the nature of the 
contaminating extended background component and only rely on the point-like 
nature of the supernova's spatial profile. 

In section \ref{algo} we present the algorithm and its application to the case
 of supernovae embedded in their host galaxy. We dedicate sections 
\ref{ssfissue} and \ref{spatialres} to the specific issues of the PSF spectrum
 -- or slit spread function (SSF) -- used to restore the point source and of 
the spatial resolution kernel used to discriminate between the point source 
and the underlying extended background, respectively, as they are of crucial 
importance to the proper functioning of \emph{specinholucy}. In sections
 \ref{testsim} and \ref{real} we test the algorithm on simulated and observed 
data, respectively. Section \ref{comparison} serves the purpose of comparing 
our technique to alternative spectral extraction methods, and we further 
discuss the advantages of using our approach alongside an optimal observing 
strategy for its successful implementation in section \ref{conclusion}.  

The need for comparisons between the high-$z$ and the local 
SN Ia spectra in order to make quantitative statements on their potential 
differences requires the use of such an elaborate two-channel deconvolution 
technique. The code \emph{specinholucy} enables a truly quantitative 
analysis of high-$z$ SN Ia spectra and opens the path for meaningful 
detections of potential evolutionary signatures in their spectra.


\section{The algorithm and its implementation \label{algo}}

	\subsection{Decomposing the SN from its host galaxy}

\citet{LW03} -- subsequently LW03 -- have described two iterative techniques 
for decomposing a long-slit spectrum into spectra of designated point
sources and an underlying background. The methods are based on a
two-channel restoration (\citealt{hooklucy94,lucy94}) using 
the Richardson-Lucy (\citealt{richardson72,lucy74}) iterative
restoration. The essence of this technique is to treat the point
source(s) and the extended background as two channels for restoration.
It must be emphasized at the outset that this restoration occurs
purely in the spatial direction and no implied spectral
restoration is undertaken. 
Thus, for instance, residuals from fringing corrections should not
be relevant unless the fringes are tilted with respect to the dispersion axis,
 in which case an additional inhomogeneous component would be added to the 
background.
The first channel contains the point 
source(s) which are restored to delta functions using an appropriate 
PSF. For a spectrum the PSF must be 
specified as a function of wavelength; this is simply the aforementioned 
SSF in \citet{LW03}.
In order to iteratively restore the extended
component, it is necessary to impose a limiting resolution, larger than
that set by the PSF, in order to prevent the second channel 
from modelling the point sources as peaks in the extended component. 
The resolution limit for the background is set by defining the 
restored background to be the convolution of a non-negative auxiliary
function with a wavelength-independent resolution kernel $R$. Then no feature
whose width is less than $R$ can appear in the restored background. 
If the width of $R$ is greater than the PSF, then the convolution of 
the model of the extended background
$\psi$ with the PSF cannot fit a point source, which must therefore
be modelled by the first channel.

Two methods are described in LW03: one for a
homogeneous background (i.e. the spectrum of the background is
homogeneous - does not vary with position); the second,
more general, case allows the spectrum of the background to vary 
as a function of position along the slit - termed the inhomogeneous 
background case. In the case of a galaxy including a SN the homogeneous 
case is appropriate.
The observed spectrum $\phi(\lambda,y)$, where $\lambda$ is the
dispersion direction and y the cross-dispersion direction, can 
then be modelled as: 

\begin{displaymath}
  \phi(\lambda,y)= \int F(\lambda,\eta) P(\lambda,y-\eta)d\eta
+ f_{\rm SN}(\lambda) P(\lambda,y-y_{\rm SN}),
\end{displaymath}

\noindent
where $\eta$ is the independent variable for the spatial
dimension (cross dispersion direction). The first term represents the spectrum
 of the background and the 
second represents the spectrum of the SN. In this study it is
$f_{\rm SN}(\lambda)$, the spectrum of the SN, which is the desired
output product. $P(\lambda,y-\eta)$ is the SSF, i.e. a PSF as a 
function of wavelength. The restored background, $\psi$, 
is a convolution of an non-negative auxilliary function 
$\chi(\lambda,\zeta)$ and the non-negative normalized resolution 
kernel centred at $\eta=\zeta$, $R(\eta-\zeta)$:

\begin{displaymath}
 F(\lambda,\eta)=\int \chi(\lambda,\zeta) R(\eta-\zeta) d\zeta,
\end{displaymath}

\noindent
which is convolved with the SSF to match the observed extended 
component. $R$ is specified as a Gaussian function
of sigma $\sigma_{\rm kernel}$.

The unknowns $f_{\rm SN}$ and $\chi$ are determined by iterative
improvement of the fit of $\phi$ to the observed spectrum, using the 
R-L algorithm to improve both the point source spectrum and the 
background until convergence is reached. The iterations steps are
described in detail in LW03. There are two convergence criteria:
the fractional change in the spectrum per R-L iteration
summed for all wavelengths; the fractional change in $\chi$ 
per iteration summed over all wavelengths and cross-dispersion range. 
In practice, depending on the particular details of the spectrum,
one of these criteria may converge much faster than the other, though 
both must be met in order for the code to converge.

	\subsection{Practical implementation for SN and host galaxy}

The situation of a SN observed in a host galaxy is an ideal
use of the inhomogeneous case (see LW03 for some other 
examples).
This is the simplest application since there is ideally
one point source and a structurally well-resolved galaxy. When the
SN is in the outer regions of an early-type galaxy, then the
sophistication of the technique is probably not essential and
more traditional methods, such as a linear or polynomial fit to
the background and extraction, with or without weighting, of 
the point source is
adequate. However, when the SN is near the centre of a galaxy,
or the galaxy does not have a smooth radial profile, or 
there is line emission in the galaxy host (e.g. for late-type 
galaxies), then the simple methods fail and
a dedicated extraction or a restoration approach is mandated. In section 
\ref{comparison} we compare the technique presented here with other less 
advanced methods.

The inhomogeneous case is implemented in the 
IRAF code \emph{specinholucy}, which is available in the 
ST-ECF layered package \emph{specres}. In addition to the input 
long-slit 2D spectrum,
there are two important input parameters: the SSF, which obviously
must match as well as possible the cross-dispersion profile 
as a function of wavelength for the SN observation and the position
of the SN on the slit. The inhomogeneous decomposition technique
requires the position of the point source that is to be extracted 
from the extended background (i.e. the host galaxy). 
Errors in this position will result in mixing of the two
channels in the output-restored spectrum so that for example 
the background is excavated asymmetrically around the point source.
However, this requirement is not as limiting as it sounds since 
one of the outputs is a 2D restored version of the input 2D long-slit spectrum 
(including both the SN and the host galaxy) which can be directly compared with
the latter spectrum and a mismatch of the position shows up as a higher 
frequency 
component than the resolution kernel in the neighbourhood of the designated SN.
The primary output is the spectrum of the
SN, which is strictly 1D. If an error image is available for the 
input long-slit spectrum, it can be supplied to produce Monte
Carlo error estimates for the restored output spectra (point source
and extended background).


\section{The importance of the slit spread function \label{ssfissue}}

LW03 discuss some of the difficulties with using an empirical
SSF. For ground-based data a spectrum of a bright standard star
is ideal, but this must be taken in identical conditions
to the SN spectrum. In practice this is almost impossible
to achieve. Even a star on the slit of the same observation as the SN cannot 
be guaranteed to provide an ideal SSF, since the star may not be centred on
the slit as the SN is centered and instrumental optical aberrations can produce
off-axis distortions of the spatial profile of a point source.
A simple task, $specpsf$, is provided to construct model SSFs in the 
$specres$ package. A number of PSFs (viz stellar
images) are provided
as input at specified wavelengths and these are sampled by a slit
of specified size (relative to the PSFs) and the signal within
a slit integrated across the slit width as a function of 
offset position (cross-dispersion). The set of point sources are
then spline interpolated in the dispersion direction to 
provide the 2D spectrum of the point source, the SSF. For an
observation with no accompanying SSF, a considerable amount of
trial-and-error is required to choose the seeing of a set of
model PSFs in order to provide a high quality restoration, and
thus successful extraction of the SN spectrum. In the low signal-to-noise
cases, it will be the major source of uncertainty on the SN
spectrum (see section \ref{simresults}). However, all methods which rely on 
provision of a PSF (in imaging) or an SSF (in spectroscopy) are subject to
this difficulty; it usually provides the fundamental limit on
the uncertainty of the restoration.

	\subsection{Measuring the seeing}

To successfully extract the point source spectrum from the extended background
 component one needs in principle to know the width of the point source 
spatial profile at each dispersion coordinate -- i.e. the FWHM-wavelength
 relation for that object. In practice however we often cannot do this due to 
poor S/N (this is especially the case in high-$z$ SN Ia spectroscopy) and so 
we rely on average quantities, such as the mean seeing during the observation.
 One can then imagine reconstructing a more reliable description of the seeing
 ($\theta$) variation along the dispersion direction ($\lambda$) using for 
instance the following seeing-wavelength relation \citep{schroeder}:

\begin{displaymath}
\left.
\begin{array}{ll}
\theta \phantom{1} =   \lambda / r_0 \\
r_0    \propto \lambda^{6/5} 
\end{array} 
\right\} \Rightarrow \theta \propto \lambda^{-0.2},
\end{displaymath}

\noindent
where $r_0$ is the so-called Fried parameter, which describes the quality of a
 wave that has propagated through atmospheric turbulence 
\citep{fried65,fried75}.

\begin{figure}
   \resizebox{\hsize}{!}{\includegraphics[width=10cm]{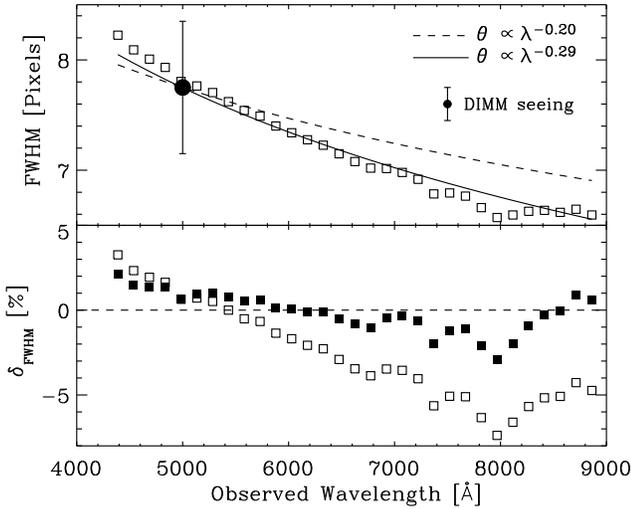}}
      \caption{\textbf{Top panel:} FWHM of the cross-dispersion profile of
 the standard star LTT 7987 taken with VLT+FORS1 on UT 11 October 2002. The 
single data point with an error bar shows the mean seeing as measured by the 
DIMM station -- namely $1.55 \pm 0.12$\arcsec\ (the pixel scale of FORS1 is 
0.2\arcsec\ pix$^{-1}$) at 5000\AA. Overplotted is the FWHM-wavelength 
relation fixed at the DIMM data point, both for the predicted exponent 
($-0.20$) and our best-fit value ($-0.29$). \textbf{Lower panel:} 
Fractional residuals with respect to the $\theta \propto \lambda^{-l}$ 
relation, for $l=0.20$ (open squares) and $l=0.29$ (filled squares).
              }
         \label{ssffwhm}
\end{figure}

However, a specific telescope and
instrument combination can introduce systematic errors due to instrumental 
spatial distortions which are not taken into account in the above relation. 
Fig. \ref{ssffwhm} shows that one should not rely on measurements by external 
seeing monitors on observatory sites -- such as the Differential Image Motion 
Monitor (DIMM) on the ESO-Paranal site, which measures the image quality via a
 differential centroiding method \citep{sarazin90,sandrock00}. Although 
precise, these measurements are inadequate for our purpose since the DIMM 
probes a region of sky significantly different from that where the telescope 
is pointing \citep[see][ Fig.~17]{patat03}. Moreover the optical train is 
different and the observation wavelength is restricted to 5000\AA. Even 
varying the exponent in the above seeing-wavelength relation does not enable 
to accurately reproduce the FWHM profile of a bright standard star (Fig. 
\ref{ssffwhm}). We will see in section \ref{simresults} that errors of 
$\gtrsim 15\%$ in the determination of the FWHM of the point source can lead 
to significant errors in the restored point source flux. One possible 
alternative is to construct a \emph{synthetic} SSF based on the 
wavelength-dependent FWHM characteristics of the point source spectrum to 
extract. This has the prime advantage of extracting the point source with a 
PSF spectrum that has been subject to the \emph{same} seeing variations. To 
do so one can either use the SN spectrum itself or carefully align the slit to
 include a bright single star as well as the object of interest.

	\subsection{Generating synthetic PSF spectra}

\begin{figure}
   \resizebox{\hsize}{!}{\includegraphics[width=10cm]{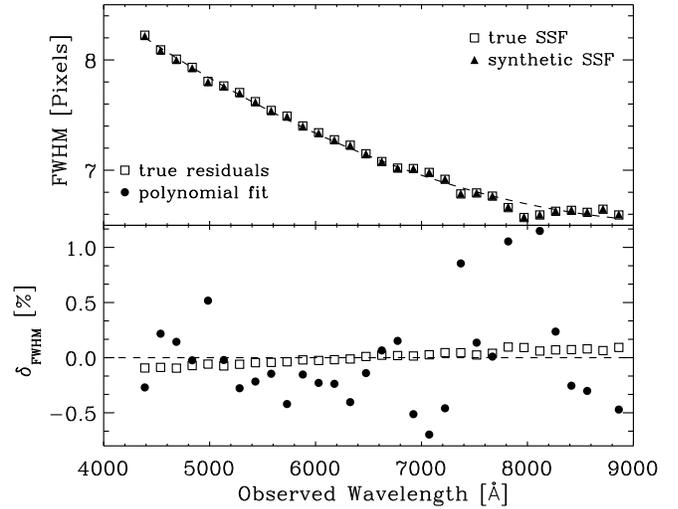}}
      \caption{Illustration of SSF synthesis using \emph{specpsf}. The input 
spectrum is the same as in Fig. \ref{ssffwhm}. \textbf{Top panel:} FWHM of
 LTT 7987 (open squares) and of the synthetic SSF (filled triangles). The 
dashed line shows a polynomial fit to the synthetic SSF. \textbf{Lower 
panel:} Residual plots of the FWHM of the synthesized SSF (open squares) and
 polynomial fit to the synthetic SSF (filled circles) with respect to the 
input SSF.  
              }
         \label{specpsf}
\end{figure}

The IRAF-implemented code \emph{specpsf} is illustrated here in Fig. 
\ref{specpsf}. A standard star is used to determine the corresponding 
FWHM-wavelength relation, from which gaussian PSF images are generated every 
50\AA. The resulting stellar images are run through \emph{specpsf} and the 
profile of the corresponding output PSF spectrum (or SSF) is compared with 
that of the input standard star. The errors fall below 0.1\% over the whole 
wavelength range. We also show a polynomial fit to the data, since in practice
 the scatter is significantly larger for low S/N point sources, and such a fit
 to the data is more representative of the actual seeing variations in the 
dispersion direction. Here the residuals are as high as $\sim$2\% in the red 
region of the spectrum (mainly due to an increase of sky brightness in this 
spectral range), still an order of magnitude below the lower threshold where 
errors on the FWHM start to matter (see below).

	\subsection{Impact of the SSF on spectral restoration 
\label{impactssf}}

The astronomer wishing to extract point source spectra using 
\emph{specinholucy} has various options for the SSF. In low S/N cases like 
the ones we describe in this paper the choice of the SSF is by far the most 
limiting factor and will have a severe impact on the quality of the 
restoration.

\begin{figure}
   \resizebox{\hsize}{!}{\includegraphics[width=10cm]{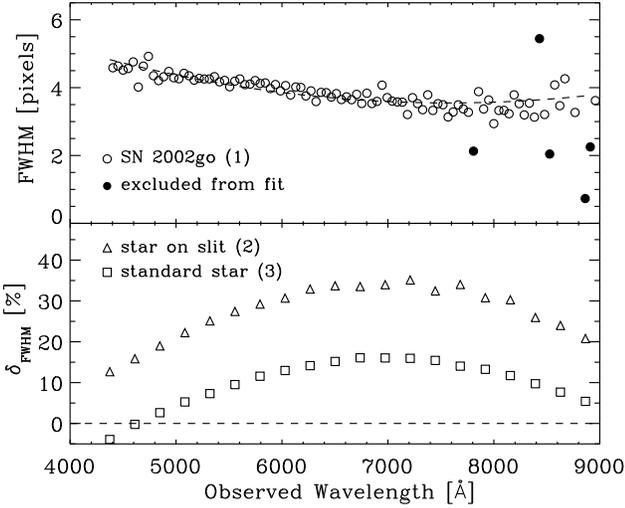}}
      \caption{Examples of SSFs used to extract SN 2002go: (1) the supernova 
itself; (2) a bright PSF star that happened to be on the same slit; (3) the 
closest FORS1 standard star to match the supernova profile. 
\textbf{Top panel:} The dashed line is a polynomial fit to the SN profile,
 from which low S/N points affected by the increase in sky background have 
been excluded. \textbf{Lower panel:} Fractional differences in the SSF 
FWHM with respect to the polynomial fit.  
              }
         \label{various_ssf}
\end{figure}

We illustrate the impact of the SSF choice on the restoration of SN 2002go, a 
Type Ia supernova at $z=0.236$ (IAUC 7994) slightly offset from the center of 
its host galaxy (see also section \ref{real}). We consider three options for 
the SSF: (1) a synthetic SSF based on FWHM measurements of SN 2002go itself, 
(2) a PSF star that happened by chance to be on the same slit as SN 2002go and
 (3) a VLT+FORS1 standard star for which the FWHM-wavelength relation was the 
closest we could find to that of SN 2002go. The resulting cross-dispersion 
profiles are shown in Fig. \ref{various_ssf}, and the corresponding mean 
seeing values are shown in Table \ref{ssffwhmdata}. Note that the quoted mean 
values are not representative of the wavelength-dependent nature of the seeing
 and one should produce plots such as those shown in Fig. \ref{various_ssf} 
to appreciate the differences between the SSFs. Despite having been observed 
on the same location of the CCD chip, the SN and the standard star were not 
observed on the same night and were thus subject to different seeing 
conditions. The difference is also a function of wavelength, hinting at 
possible instrumental effects on the spatial profile. These same comments 
apply to standard stars having been observed on the same night as the object 
of interest, and we discourage the potential user of this method from using 
standard stars for the SSF.

\begin{table}
\caption{SSF FWHM data}
\label{ssffwhmdata}
\centering
\begin{tabular}{l l l}
\hline\hline
SSF	& pixels$^{a}$  & arcseconds$^{a}$  \\
\hline
SN 2002go 	& $3.83^{4.74}_{3.48} \pm 0.41$	& $0.77^{0.95}_{0.70} \pm 0.08$ \\
star on slit	& $4.71^{5.44}_{4.18} \pm 0.39$ & $0.94^{1.10}_{0.84} \pm 0.08$ \\
standard star	& $4.17^{4.64}_{3.93} \pm 0.23$ & $0.83^{0.93}_{0.79} \pm 0.05$ \\
\hline
\end{tabular}
{\footnotesize
\flushleft
(a) the quoted values are read as 
$<{\rm mean}>^{\rm max}_{\rm min} \pm 1\sigma$ error
}
\end{table}

More interesting, perhaps, are the quasi-systematic differences in
the profiles of the SN and the PSF star that was accidentally placed
on the same slit.  The SN and PSF star should be subject to identical
seeing variations, but the reconstructed profiles differ by as much as
$\sim30\%$, or $0.2$\arcsec. There could be several reasons for this: (a) the 
observation is not made at the parallactic angle, and so the dispersion 
direction is not along the slit; (b) the star spectrum is located at a 
different location on the CCD chip where the spatial distortion differs; (c) 
the star is not centred on the slit and we are not measuring its true PSF 
profile. FORS1 is equipped with a longitudinal atmospheric dispersion 
compensator (LADC)
which has an accuracy $\lesssim 0.1\arcsec$ (0.5 pix) over the considered 
wavelength 
range \citep{avila97}, and so (a) should not be relevant. FORS1 is linear at 
the $<0.4\%$ level (see the FORS1+2 User Manual
\footnote{http://www.eso.org/instruments/fors1/}) and distortions associated 
with the instrument's optics should not affect the 
whole spectral range at this level, so (b) should not matter either. This 
leaves (c) as the probable explanation for this difference, which is 
significant enough to be a major source of error in the restoration.

\begin{figure}
   \resizebox{\hsize}{!}{\includegraphics[width=10cm]{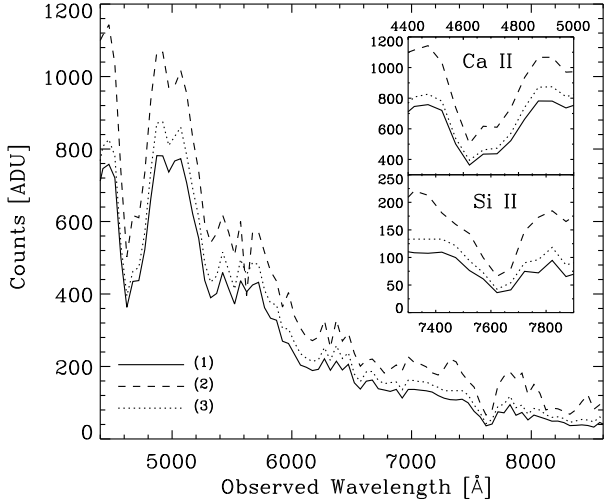}}
      \caption{Impact of the SSF on the restored point source spectrum of SN 
2002go. Refer to Fig. \ref{various_ssf} for the meaning of the line 
annotations (1), (2) and (3), noting that (1) corresponds to the polynomial 
fit to the SN 2002go profile. The two inlays zoom in on the blueshifted Ca II 
H\&K and Si II features, both prominent in SN Ia spectra, and highlight the 
impact of non-optimal spectral extractions on empirical correlations involving
 line strengths (see text).
             }
         \label{ssfimpact}
\end{figure}

The impact of these various SSFs on the restored SN flux is shown in Fig. 
\ref{ssfimpact}. We view the solid curve as the most accurate restoration, 
and one sees that using the other two (wider) SSFs means we are also restoring
 part of the extended background component (sky and host galaxy) in the point 
source spectrum. In the case of the star that was by chance located on the 
same slit the 
$\sim 20\%$ difference in the FWHM results in an error in the restored SN 
flux that can be as high as 100-150\%! We see from 
the two inlays in Fig. \ref{ssfimpact} that the slope of the 
pseudo-continuum around strong SN Ia spectral features (Ca II H\&K and Si II) 
changes significantly due to contamination of the underlying background 
spectrum, and so the very definition of ``line strength'' in this case is very 
ambiguous. The fact that the FWHM of the point source varies with wavelength 
will cause variable contributions of the background to the restored spectrum, 
and will affect the determination of line ratios of distantly spaced lines, 
if the PSF vector is not determined properly.
Thus, the extraction has a significant impact on empirical correlations based 
on such measurements, as the errors made at this stage will add to those 
associated with the subsequent calibration in flux. Examples of such 
correlations are given in \citealt{nugent95} and are used to determine 
distances to SNe Ia in \citealt{riesssnapshot}.

One might ask how one can decide which SSF to choose to restore a particular 
point source spectrum and if it is possible to \emph{a posteriori} tell 
whether the chosen SSF was indeed the appropriate one. Indeed, through 
restoring the complete 2D background spectrum \emph{specinholucy} enables 
us to check the accuracy of the restoration at each pixel of the input 2D 
frame, or simply in the spatial (cross-dispersion) direction by comparing the 
collapsed spatial profiles of the input and restored 2D spectra. This is 
illustrated using real data in Fig. \ref{specinhofig1}.

	\section{The spatial resolution kernel \label{spatialres}}

\begin{figure}
   \resizebox{\hsize}{!}{\includegraphics[width=10cm]{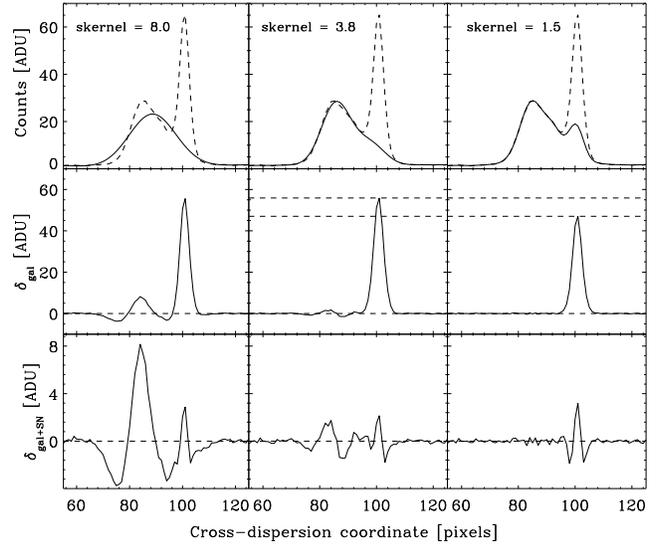}}
      \caption{Impact of the width of the spatial resolution kernel 
$\sigma_{\rm kernel}$ on the background restoration. \textbf{Top panel:} 
Input average (dashed line) and restored background (solid line) 
cross-dispersion profiles. \textbf{Middle panel:} Residuals of the 
previous plot, revealing the point source channel (i.e. the supernova itself).
 \textbf{Lower panel:} Spatial residuals of the input and restored 2D 
frames. Negative (positive) values in the residual plots mean we are over
(under)-restoring the flux at that location.
              }
         \label{skernel}
\end{figure}

The choice of the width of the resolution kernel for the background resides 
with the user. It is implemented as a Gaussian of user-chosen sigma 
$\sigma_{\rm kernel}$ in the \emph{specinholucy} code. It is difficult to 
give a general guide to its choice. Obviously it should be wider than the 
equivalent maximum Gaussian sigma of the SSF $\sigma_{\rm SSF,max}$ in the 
cross-dispersion direction.
The sigma of the kernel has to be tuned and is very dependent 
on the nature of the 
extended source and more specifically on the extended background component in 
the vicinity of the point source. 
Comparison of the restored background with the input spectrum will
quickly show if narrow features in the extended background, above the noise, 
have failed to be modelled through choice of too wide a kernel width (see Fig.
 \ref{skernel}, left column). If the kernel width is too narrow then the point
 source will be partially modelled as a peak in the background channel and the
 fluxes in our restored point source spectrum will be underestimated (Fig. 
\ref{skernel}, right column). The optimal width for the spatial resolution 
kernel corresponds to a compromise between the lower spatial frequency of the 
extended background cross-dispersion profile and the higher spatial frequency 
of the point source cross-dispersion profile. A decent first guess would be 
$\sigma_{\rm kernel} \approx 2$-$3\overline{\sigma_{\rm SSF}}$. In Fig. 
\ref{skernel} (middle column) we have used 
$\sigma_{\rm kernel} \approx 2.3\overline{\sigma_{\rm SSF}} > 
\sigma_{\rm SSF,max}$. From Fig. \ref{skernel} we further note that, if one 
can clearly see the impact of a wide kernel on the spatial residuals 
$\delta_{\rm gal+SN}$, discriminating between the narrow and optimal kernel 
widths -- respectively 1.5 and 3.8 -- is not obvious \emph{a priori}. The 
optimal kernel width in Fig. \ref{skernel} was chosen such as to maximise the 
counts in the point source channel, or $\delta_{\rm gal}$, whilst keeping the 
spatial residuals $\delta_{\rm gal+SN}$ below the statistical noise limit of 
the input 2D frame.

One important test is to ensure we are not capable of restoring a PSF-like 
spectrum from a pure extended source. This is particularly relevant to 
high-$z$ SN Ia spectroscopy since in many cases one is only able to extract a 
spectrum of the supernova host galaxy whereas one is convinced of having 
obtained a combined spectrum including the SN. Then the currently widespread 
technique of subtracting a galaxy template from this ``combined'' (and noisy) 
spectrum can sometimes reveal SN-like features where no supernova is present! 
This is illustrated later on in section \ref{snfit} where we compare the 
\emph{specinholucy} output with that of a statistical algorithm 
($\mathcal{SN}$-fit) which decomposes a 1D spectrum into its SN and galaxy 
components using a galaxy spectral template (see also Fig. \ref{compare}).

\begin{figure}
   \resizebox{\hsize}{!}{\includegraphics[width=10cm]{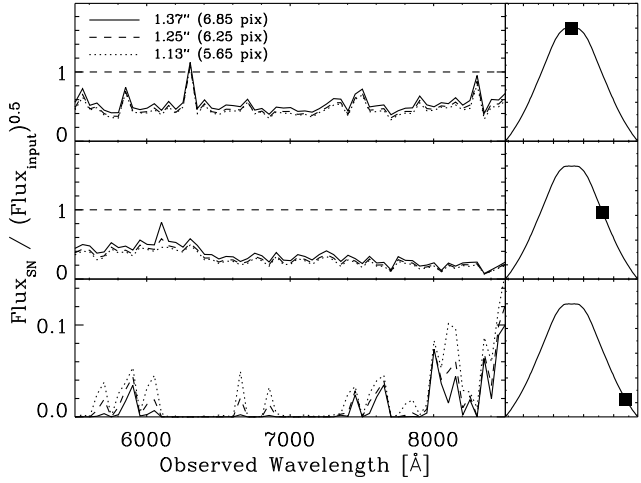}}
      \caption{Restored point source spectra in units of the statistical noise
 of the input pure galaxy spectrum. The different plots correspond to 
different positions of the point source to extract, marked by a filled square 
in the galaxy spatial profiles shown in the right column. The different lines 
correspond to different seeing conditions. The horizontal dashed line marks 
the limit below which no point source was detected in the background. Note the
 change of scale in the lower plot, where no flux at all is restored in many 
bandpasses. The peak above the dashed line in the uppermost plot is due to bad
 cosmic ray removal.
              }
         \label{failedex}
\end{figure}

Here we have used the uncontaminated galaxy spectrum of the host of SN 2002bo 
(see section \ref{real}), obtained by constructing a mirror image of the 
portion of the galaxy profile devoid of supernova signal. We try to extract a 
point source component from this pure galaxy spectrum at different locations 
of the galaxy trace, including the centroid, using for this purpose three 
different SSFs of widths corresponding to the mean and extrema seeing values 
measured by the DIMM station for this observation. We then compare the 
restored point source flux with the statistical noise limit of the input 2D 
spectrum at the position of the point source. Should this ratio fall below one
 this means that no point source was detected. We see from Fig. \ref{failedex}
 that in all cases the restored spectrum is inconsistent with that of a 
PSF-like object over the whole spectral range.

\section{Testing the algorithm on simulated data \label{testsim}}

In this section we present tests of the algorithm on simulated data, namely 2D
 spectra of a Type Ia supernova (SN Ia) at $z=0.5$ embedded in a late-type 
galaxy. We have chosen to reproduce a situation where a one hour-long exposure
 is taken at the ESO Very Large Telescope VLT-UT1 (8.2m) using the FOcal 
Reducer Spectrograph (FORS1) in Long-Slit Spectroscopy (LSS) mode. The grism 
corresponds to 300V, and the slit width is 1\arcsec. The reason behind chosing
 these specific settings is that they correspond to those used for the real 
data on which the algorithm will be tested further in section \ref{real}. In 
what follows we assume a $(\Omega_M,\Omega_\Lambda,h)=(0.3,0.7,0.72)$ 
cosmology (where $h=\frac{H_0}{100\ {\rm km\ s}^{-1} {\rm Mpc}^{-1}}$ is the 
dimensionless Hubble constant), and so the luminosity distance corresponding 
to $z=0.5$ is $d_{{\rm L,}z=0.5} \sim 2.72$Gpc. The factors used to scale down
 the fluxes as the SN and galaxy are artificially redshifted to $z=0.5$ are 
listed in Table \ref{distances}. The additional factor of $(1+z)$ comes from 
the fact that we are integrating the flux over a finite bandpass, where the 
wavelength axis has been diluted by that same factor.

\begin{table}
\caption{Distances and flux scaling}
\label{distances}
\centering
\begin{tabular}{l l l l}
\hline\hline
galaxy 		& d (Mpc)& Reference 		      & factor$^{\rm a}$     \\	
\hline						                             
NGC 4526	& 13	 & \citealt{ngc4526}          & $1.5 \times 10^{-5}$ \\
NGC 6181	& 34	 & \citealt{silchenko97}      & $1.1 \times 10^{-4}$ \\
\hline
\end{tabular}
{\footnotesize
\flushleft
(a) Factor used to scale down the flux, computed as 
$(d/d_{{\rm L,}z=0.5})^2 \times (1+z)^{-1}$. \\
}
\end{table}

	\subsection{Simulated data}

The simulated 2D spectra are a combination of a 2D supernova spectrum and a 2D
 background spectrum, itself consisting of a galaxy and a sky spectrum. We 
vary the phase of the SN, i.e. the brightness and the input spectrum, and its 
position within the galaxy from 0.75 to 1.75, in units of FWHM of the galaxy 
trace. Random poisson noise is added to the image using the gain and read 
noise characteristics of FORS1 so as to degrade the overall S/N of the image 
and thereby reproduce plausible observing conditions.

		\subsubsection{Supernova spectrum}

The 2D supernova spectra are synthesized using the 1D, flux-calibrated spectra
 of the Type Ia supernova SN 1994D in NGC 4526 ($cz=448$ km s$^{-1}$) as 
obtained from the SUSPECT 
database\footnote{http://bruford.nhn.ou.edu/$\sim$suspect/.}. Where spectral 
coverage at a given phase was lacking (i.e. at $-6$d, $+0$d, $+6$d, $+8$d and 
$+14$d), approximate synthetic spectra were obtained from the two observed 
spectra closest in phase using the $UBVRI$ photometry published in 
\citet{patat94D}. The same technique was applied to observed spectra whose
 wavelength range did not extend below 3650\AA\ in the blue 
(at $-8$d and $+10$d), so as to make sure to cover the Ca II H\&K 
(3934\AA,3968\AA) absorption trough blueshifted to  $\sim$3750\AA, prominent 
in SNe Ia optical spectra (see \citealt{filippenko97} for a review). 
Each spectrum is then redshifted to $z=0.5$ and rebinned to the pixel scale 
of FORS1 equipped with a 300V grism and a 1\arcsec slit 
($\sim 2.66$\AA\ pix$^{-1}$). The flux (in units of erg s$^{-1}$ 
cm$^{-2}$ \AA$^{-1}$) is scaled down to account for the change in luminosity 
distance as the source is artificially placed at $z=0.5$ -- assuming SN 1994D 
to be at the same distance as its host galaxy -- and is converted to counts 
(ADU) using a FORS1 sensitivity function. At this stage we have a 1D supernova
 spectrum in ADUs as a function of wavelength, corresponding to a one-hour 
integration on FORS1. Finally, a 2D supernova spectrum is synthesized by 
replicating this 1D spectrum along the lines of a 2D image array (i.e. along 
the dispersion axis) and multiplying each image column by a 1D gaussian 
profile normalised to unity of FWHM corresponding to 1\arcsec seeing. Note 
that the flux vector is normalised to an airmass of 1, and no scaling was 
applied to account for observations (usually) made at higher airmasses.

 The supernova data used in this simulation are summarised in Table 
\ref{sndata}. The quoted $R$-band magnitudes are only approximate and are 
affected by systematic effects due to the lack of any $k$-correction. For 
comparison the mean number of counts (in ADUs) per pixel is 105.7 and 19.8 for
 galaxy and sky (including sky lines), respectively. 

\begin{table}
\caption{SN 1994D data used in the simulation}
\label{sndata}
\centering
\begin{tabular}{l c l c c}
\hline\hline
phase$^{\rm a}$  & date$^{\rm b}$ & $m_B\ ^{\rm c}$ & $m_{R,z=0.5}$$^{\rm d}$ & counts$^{\rm e}$ \\
\hline
$-$10  & 11/03/94  & 13.26  & 24.86  & 13.5     \\
$-$8   & 13/03/94  & 12.67  & 24.27  & 22.7	\\
$-$6   & $\cdots$  & 12.26  & 23.86  & 30.7	\\
$-$4   & 17/03/94  & 11.99  & 23.59  & 37.4	\\
$-$2   & 19/03/94  & 11.86  & 23.46  & 39.5	\\
+0     & $\cdots$  & 11.84$^{\dag}$  & 23.44  & 51.6	\\
+2     & 23/03/94  & 11.89  & 23.49  & 48.4	\\
+4     & 25/03/94  & 11.99  & 23.59  & 34.8	\\
+6     & $\cdots$  & 12.13  & 23.73  & 31.3	\\
+8     & $\cdots$  & 12.33  & 23.93  & 27.9	\\
+10    & 31/03/94  & 12.55$^{\dag}$  & 24.15  & 21.5	\\	
+12    & 02/04/94  & 12.80  & 24.40  & 15.0	\\
+14    & $\cdots$  & 13.10  & 24.70  & 12.7	\\
\hline
\end{tabular}
{\footnotesize
\flushleft
(a) SN phase in rest-frame days from $B$-band maximum. \\
(b) UT date of spectroscopic observation. Absence of date means the spectrum 
is synthetic. \\
(c) $B$-band magnitude taken from Table 1 in \citet{patat94D}. Data marked 
with $^{\dag}$ results from a polynomial fit to the $B$-band light curve. \\
(d) the $R$-band at $z=0.5$ is close to rest-frame $B$-band, and so $m_R\ 
(z=0.5)\approx m_B - 2.5\log{\left( \frac{d_{L,SN}}{d_{L,{\rm gal}}} \right) 
^2} \approx m_B + 11.60$. \\ 
(e) mean total counts in the whole SN profile in ADUs per pixel along the 
dispersion axis. \\
}
\end{table}

		\subsubsection{Background spectrum}

\begin{figure}
   \resizebox{\hsize}{!}{\includegraphics[width=10cm]{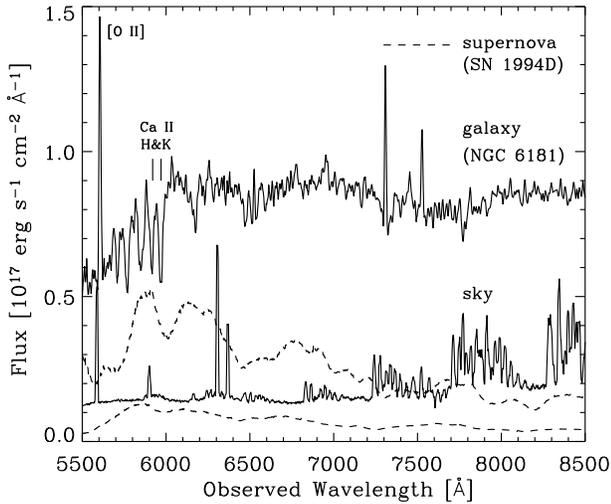}}
      \caption{Absolute flux-calibrated galaxy and sky spectra used in the 
simulation. The spectra are shown in observed wavelength at $z=0.5$, and 
without shot noise for sake of clarity. The prominent [O II] emission and the 
Ca II H\&K absorption in NGC 6181 are labelled accordingly. Overplotted 
(dashed lines) are the spectra of SN 1994D at maximum intensity (top) and ten 
days before maximum (bottom).
              }
         \label{backspec}
\end{figure}

For the galaxy spectrum we choose to artificially place SN 1994D in NGC 6181, 
an SC galaxy part of Kennicutt's spectrophotometric atlas of galaxies 
\citep{kenn1} . The reason behind this is the prominent [O II] emission at 
3727\AA\ present in its spectrum (see Fig. \ref{backspec}), which could then 
serve as an diagnostic tool for probing the efficiency of our algorithm in 
restoring a clean point source spectrum. If successful, the restored 1D 
spectrum of the point source should be devoid of [O II] flux residuals. 

The spectrum of NGC 6181 was redshifted to $z=0.5$ and rebinned to the pixel 
scale of FORS1 using the same procedure as for the supernova spectrum. The 
flux was scaled down to account for the change in luminosity distance, and 
converted to counts (again in ADUs) using the same sensitivity function as for
 the supernova. The resulting galaxy has a magnitude $m_{\rm gal}\sim21.5$ in 
a reshifted $B$-band filter (very roughly corresponding to the $R$-band at 
$z=0.5$) i.e. more than an order of magnitude brighter than when the supernova
 is at its faintest (see Table \ref{sndata}).

Next, a galaxy spatial luminosity profile was constructed based on a standard 
bulge$+$disk model. The bulge component is usually described analytically 
using the so-called $r^{1/4}$ law and the disk component follows an 
exponential decline with distance $r$ from the nucleus 
\citep{deVaucouleurs,freeman70}. One can derive a global analytical form of 
the surface brightness profile $\Sigma_S(r)$ of spiral galaxies, as a function
 of the effective radius $r_e$ of the bulge only. In doing so we assume a 
bulge-to-total luminosity ratio $B/T = 0.4$ \citep{Ratnatunga} and a ratio 
between the effective radii of the bulge and disk components 
$r_{b,e}/r_{d,e}=0.5$ \citep{kent85,Ratnatunga}:

\begin{eqnarray*}
	\Sigma_S(r) & = & \Sigma_b(r) + \Sigma_d(r) \\
	            & = & 0.76931\ \Sigma_{S,e} \exp \left(-7.6692 \left[ 
\left({\frac{1.6617r}{r_e}}\right)^{1/4} - 1\right]\right) \nonumber \\
                    &   & + 2.9343\ \Sigma_{S,e} \exp 
\left(-\frac{1.3945r}{r_e}\right),
\end{eqnarray*}

\noindent
where $\Sigma_b(r)$ and $\Sigma_d(r)$ are the surface brightness profiles of 
the bulge and disk components, respectively, and $\Sigma_{S,e}(r)$ is the 
surface brightness at the effective radius $r_e$. 

\begin{figure}
   \resizebox{\hsize}{!}{\includegraphics[width=10cm]{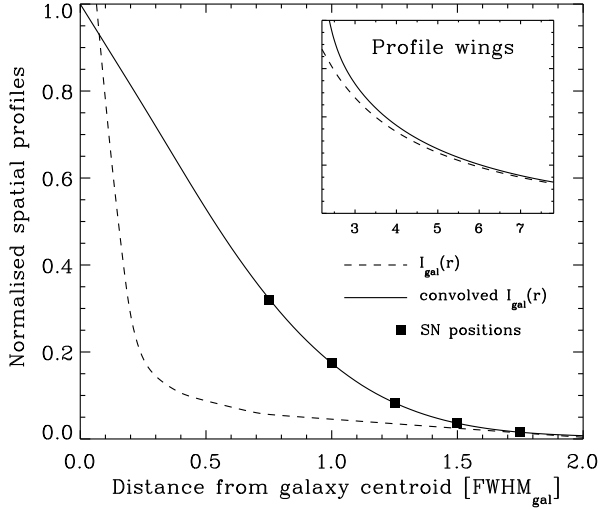}}
      \caption{Smoothed galaxy spatial profile used in the simulation, 
resulting from the convolution of a standard spiral galaxy luminosity profile 
and a 1\arcsec\ seeing gaussian kernel. The abscissa is in units of FWHM of 
the convolved galaxy profile ($\cong3.06$ pix). The effective radius $r_e$ in 
these units is at $\sim7.36$. Filled squares indicate the position of the SN 
used in this simulation, namely 0.75, 1.00, 1.25, 1.50 and 1.75. 
              }
         \label{galprofile}
\end{figure}

The angular size of NGC 6181 is 2.5\arcmin$\times$1.1\arcmin\ (NASA/IPAC 
Extragalactic Database\footnote{http://nedwww.ipac.caltech.edu/}), which 
corresponds to 4.5\arcsec$\times$2.2\arcsec\ at $z=0.5$ , and in turn to 
22.5$\times$11 pixels on a FORS1 image. Assuming the slit was placed along 
the major axis of NGC 6181 we can assume that the effective radius of the 
disk $r_e$ is 22.5 pixels. The derived luminosity profile is then normalised 
to unity and is multiplied into each column of the 2D galaxy image array. Each
 column is further convolved with a gaussian profile of FWHM corresponding to 
1\arcsec\ seeing to reproduce similar observing conditions as for the 
supernova. The resulting spatial profile $I_{\rm gal}(r)$ is shown in Fig. 
\ref{galprofile}. Notice how at these redshifts the core spatial profile of 
galaxies is entirely seeing-dominated, and one must rely on the broad extended
 wings of the profile to differentiate between a galaxy and a point source.

The sky 2D spectrum (the second background component for ground-based 
observations) is generated by uniformly replicating an observed 1D sky 
spectrum with the required settings (300V grism, 1 \arcsec\ slit) along lines 
of an image array, and scaled to the required exposure time (3600s in our 
case). It is thus by construction perfectly flat and monochromatic in the 
spatial direction. We do not have to worry about spatial distortions of sky 
lines and sky background removal in the input 2D spectrum is rendered trivial.

	\subsection{Simulation steps \label{simsteps}}

We now move on to outlining the steps made in carrying out the simulation. Far
 from being redundant these are the same steps that one should in principle go
 through when applying this technique to real data. They include determining 
the position (and slope) of the supernova spectral trace, its FWHM as a 
function of wavelength, synthesizing a corresponding SSF and running the 
\emph{specinholucy} code with the optimal settings. For the purpose of this 
simulation and for ease in computing statistics these steps have been made 
automatic, though we strongly advise to make them highly interactive when 
dealing with real data.

		\subsubsection{SN trace position and slope \label{pos}}

One of the inputs to the \emph{specinholucy} routine is a position table 
containing the $(x,y,$slope pix$^{-1}$) of the supernova spectral trace. Since
 in the case of high-$z$ SNe Ia we are limited in S/N we cannot simply give an
 approximate position and rely on cross-correlations with the SSF to determine
 a more accurate one. Here we must input the \emph{exact} SN trace position 
to which the SSF profiles will be shifted. The results of our simulation (see 
section \ref{simresults}) show the impact of positional accuracy on the point 
source restoration.

To some extent one should base the method used to determine the trace position
 on the characteristics of the instrument used for the observation. In the 
case of FORS1 we know from observations of standard stars that the position of
 a point source does not vary by more than one pixel across the whole CCD 
chip. Thus, a $\sim 1 {\rm pix}$ accuracy can almost routinely be achieved by 
summing up each of the spatial channels and finding the maximum intensity 
(other than the galaxy) in a given cross-dispersion region. Clearly this will 
be problematic for cases where the SN is close to the centroid of the galaxy 
trace or when its phase is far from maximum ($-$10d or +14d). Alternatively 
one can try to fit the background around the SN trace and subtract this fit 
from the overall profile. The residuals can then be fit with a gaussian 
profile and the trace position is taken to be the profile centre. 

\begin{figure}
   \resizebox{\hsize}{!}{\includegraphics[width=10cm]{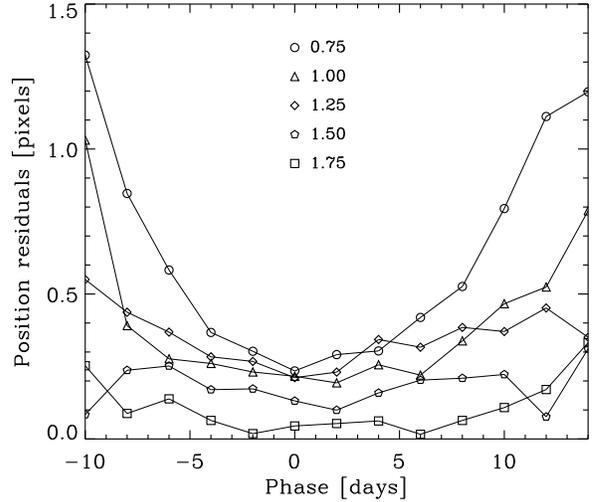}}
      \caption{SN trace position residuals as a function of the SN phase, for 
different locations of the SN with respect to the galaxy centroid. Each curve 
corresponds to a specific SN position, given in units of FWHM$_{\rm gal}$.
              }
         \label{poserr}
\end{figure}

The result from this operation is a relation between SN trace position and 
dispersion coordinate $x$, which is fit linearly so as to determine the trace 
position and slope. We now have our input $(x,y,$slope pix$^{-1}$) table 
necessary to run \emph{specinholucy}. A plot of the error made in 
determining the trace position using this approach is shown in Fig. 
\ref{poserr}. The position is best determined when the signal-to-noise ratio 
is high, namely when the contrast is high between the SN and the underlying 
background , as expected.

		\subsubsection{SN trace FWHM and SSF synthesis \label{fwhm}}

\begin{figure}
   \resizebox{\hsize}{!}{\includegraphics[width=10cm]{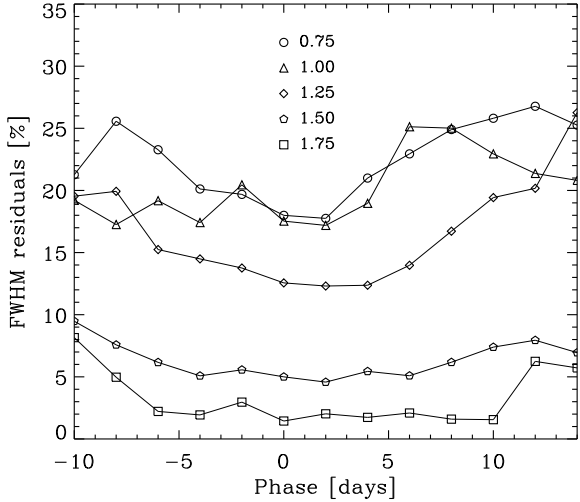}}
      \caption{SN trace FWHM residuals as a function of the SN phase, for 
different locations of the SN with respect to the galaxy centroid. The curves 
and plotting symbols have the same meaning as in Fig. \ref{poserr}.
              }
         \label{fwhmerr}
\end{figure}

To construct a synthetic SSF we need to know the FWHM of the SN trace as a 
function of the dipersion coordinate. We proceed in a way analogous to 
determining the SN trace position, except we are now interested in the width 
of the gaussian fit to the residual signal left over after sky and galaxy 
subtraction rather than the gaussian centre. A similar relation between 
FWHM$_{\rm SN}$ and dispersion coordinate is established and a table of 
(FWHM$_{\rm SN}$,wavelength) values is elaborated and serves as input for 
the SSF synthesis using \emph{specpsf}. A plot of the error made in 
determining the trace FWHM using this approach is shown in Fig. \ref{fwhmerr}.
 Again, the FWHM is better determined for cases with a higher SN flux relative
 to the galaxy background.

		\subsubsection{Running \emph{specinholucy} 
\label{runningspecinholucy}}

Once we have determined the position (and slope) of the SN trace and 
synthesized an SSF spectrum using its FWHM-wavelength dependency, we can run 
\emph{specinholucy} on the input sky-subtracted 2D spectrum. We refer the 
reader to \citet{LW03} and to the IRAF help pages for the 
\emph{specinholucy} task for a more complete description of the input 
parameters. A centroid fitting is used to determine the SSF peak at each 
column, and no subsampling (in the spatial direction) is performed since the 
resulting PSF profile is not oversampled with respect to the input 2D 
spectrum. A cubic spline is used as the interpolation method to shift the SSF 
to the position of the point source -- which is defined \emph{exactly} in 
the input position table. Note that no input statistical image to determine 
the statistical error at each pixel is used in these simulations.
	
The only input parameter made to vary from one run to the other is the width 
$\sigma_{\rm kernel}$ of the gaussian used for the spatial resolution kernel. 
We see from Fig. \ref{galprofile} that the galaxy spatial profile is entirely 
seeing-dominated, meaning that the widths of the SN and galaxy traces are 
comparable at each wavelength. This in turn implies that the spatial frequency
 of the galaxy profile will vary hugely when varying the SN position from 
0.75 to 1.75 FWHM$_{\rm gal}$. To avoid confusion in the point/extended source
 discrimination we must use a narrow (wide) spatial resolution kernel when 
the SN is close to (far from) the galaxy centroid, where the spatial frequency
 is high (low).The input values for the smoothing kernel were chosen such as 
to minimise the mean flux residuals in restoring all phases of the SN spectrum
 at a given position. For all runs the convergence criteria for the SN and 
background spectra per R-L iteration (see section \ref{algo}) were set to 
0.1\% and 1\%, respectively.

Two runs (A \& B) were executed: in run A we use the SN position and FWHM as 
determined using the method outlined in sections \ref{pos} and \ref{fwhm}, 
whilst in run B we use the known values of the SN trace position and FWHM to 
synthesize the SSF and restore the SN spectrum. This enables us to investigate
 the joint impact of the errors on the SN position and FWHM on the accuracy of
 the SN spectral restoration (see section \ref{simresults}).

		\subsubsection{Statistical calculation \label{statcalc}}

To evaluate the efficiency of the method we compare the (known) input SN 
spectrum with its restored version. More specifically we compute the ratio of 
the flux residuals with the statistical variation in the input 2D spectrum. 
Should this ratio fall below one this means that we are statistical 
noise-limited and cannot improve on the restoration.
	
In this simulation we compare the flux residuals both over the whole observed 
spectral range (5500\AA$-$8500\AA) and around the region of host galaxy 
[O II] 3727\AA\ emission. This region is simply defined in wavelength space 
(in \AA) as $3727(1+z) \pm {\rm FWHM_{[O\ II]}}(1+z)$, where 
FWHM$_{\rm [O\ II]}$ is the width of the [O II] line in the input 
non-redshifted spectrum of NGC 6181 and $z=0.5$ is the redshift of the 
simulation. For both cases the flux residual $\delta F$ is evaluated at each 
pixel according to:

\begin{displaymath}
\delta F = \frac{ \overline{ |F_{\rm SN,i} - F_{\rm SN,r} }| } 
{ \sqrt{ F_{\rm SN,i} + F_{\rm gal,i} + F_{\rm sky,i} }},
\end{displaymath}
	
\noindent
where $F_{\rm SN,i}$ and $F_{\rm SN,r}$ are the input and restored SN fluxes, 
respectively. $F_{\rm gal,i}$ and  $F_{\rm sky,i}$ are the galaxy and sky 
fluxes at the location of the SN, defined as an interval centred on the exact 
SN trace position and of width the FWHM of the SN trace. It is important to 
compute the statistical noise as that in the \emph{total} signal at the
 position of the supernova (i.e. including the underlying background) and not 
just in the SN signal itself. If $\delta F \leq 1$ one can in principle 
consider the SN restoration to have converged.

	\subsection{Simulation results and discussion \label{simresults}}

\begin{figure*}
\centering   	
   \includegraphics[width=17cm]{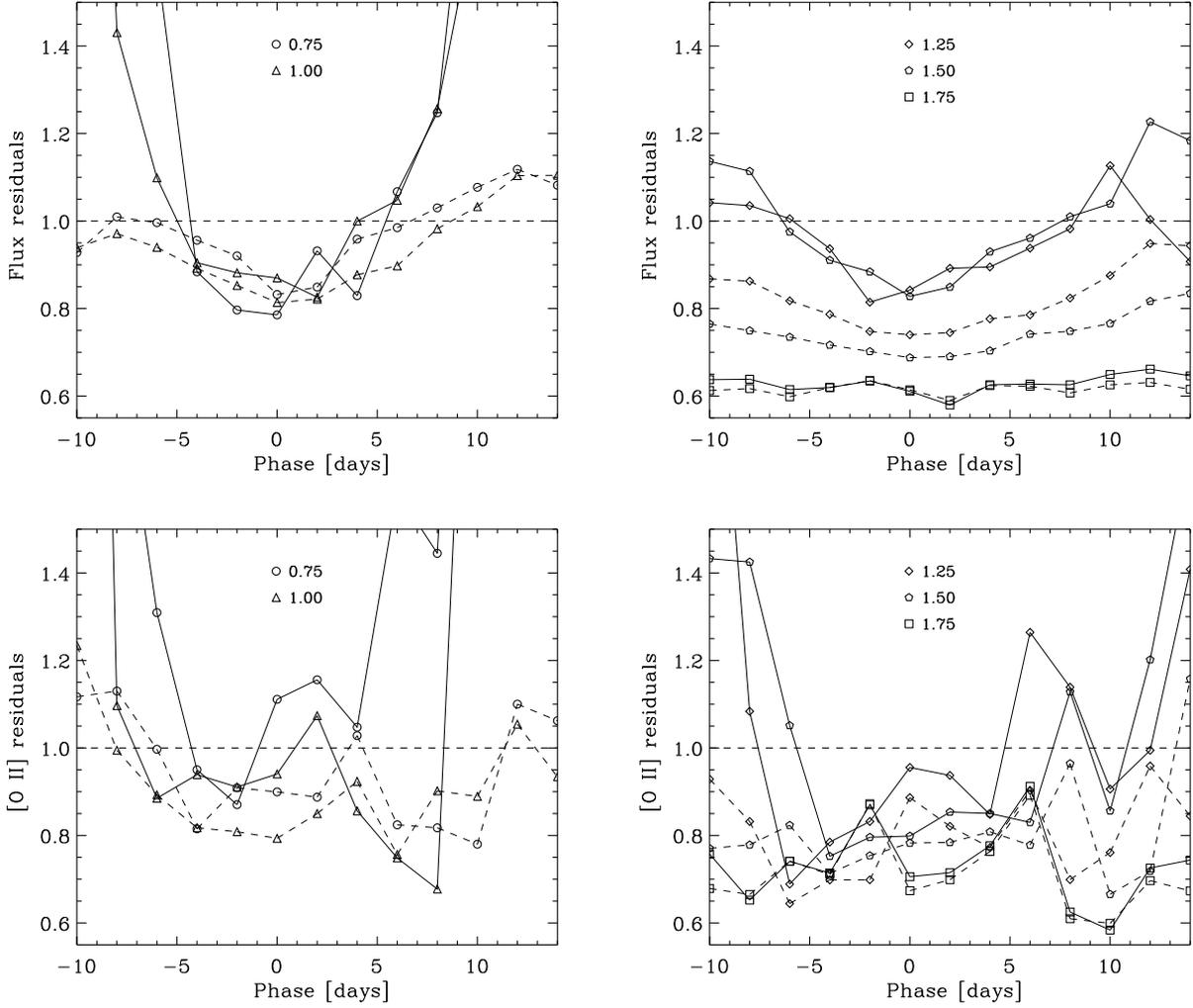}
   \caption{\textbf{Top panel:} Flux residuals $\delta F$ as a function of
 the SN phase for different positions of the SN with respect to the centroid 
of the galaxy trace, given in units of FWHM$_{\rm gal}$. $\delta F \leq 1$ 
means the SN spectrum has been restored to the statistical noise limit. The 
solid and dashed lines correspond to runs A and B, respectively (see section 
\ref{runningspecinholucy}). \textbf{Lower panel:} [O II] flux residuals as
 a function of phase for different SN positions. 
              }
   \label{simres}
\end{figure*}

The simulation results are summarised in Fig. \ref{simres}. Both runs A and B 
(see section \ref{runningspecinholucy}) are shown in order to evaluate the 
impact of the error made on the determination of the SN trace position and 
FWHM on the restoration of the supernova spectrum.

These plots clearly show that the limiting factor of this two-channel 
restoration technique (as for any other spectral extraction method) is the 
contrast between the point source and the underlying background. The overall 
flux is systematically restored to the statistical noise limit when the 
supernova is at its brightest stages 
($-5\rm d \lesssim {\rm phase} \lesssim +5\rm d$), irrespective of its 
position with respect to the centroid of the galaxy trace. In this phase 
regime the error on the SN trace position and FWHM is also at its lowest 
(see Figs. \ref{poserr} and \ref{fwhmerr}) and has little impact on the 
quality of the restoration -- as shown by the dashed lines in Fig. 
\ref{simres}. This is not always the case for the [O II] flux residuals, 
where the added background noise due to the presence of this emission line 
affects the restoration (cf. at positions 0.75 and 1.00, in units of 
FWHM$_{\rm gal}$). 

One sees the dramatic impact of the error made on the SN trace position and 
FWHM at low S/N, namely when the SN is outside the 
$-5\rm d \lesssim {\rm phase} \lesssim +5\rm d$ range and close to the 
centroid of the galaxy (at positions of 0.75, 1.00 and 1.25). For errors in 
the position $\gtrsim0.3\ {\rm pix}$ and FWHM$_{\rm SN} \gtrsim 15\%$, the 
restoration fails to reach the statistical noise limit. These are the main 
sources of systematic error in the restoration method and great care should be
 taken when determining the SN trace position and FWHM when applying it to 
real data.


\section{Testing the algorithm on real data \label{real}}

\begin{figure*}
\centering   	
   \includegraphics[width=18cm]{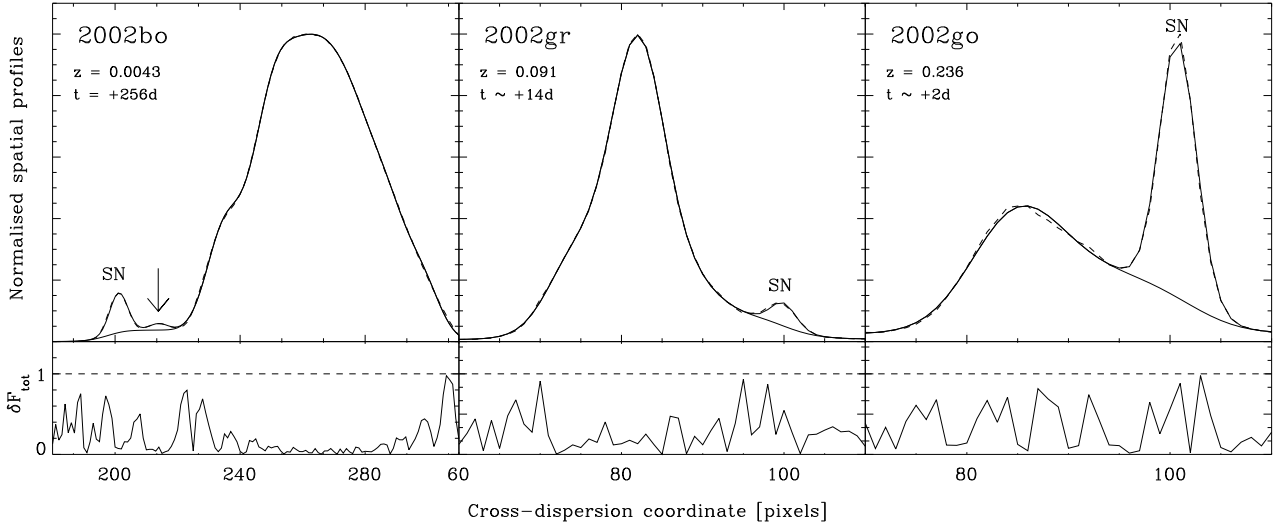}
   \caption{\textbf{Top panel:} Normalised average spatial profiles of the
 input (dashed line) and restored (solid line) 2D background spectra. Two runs
 of \emph{specinholucy} were executed, one including the point source in the
 restored 2D background spectrum and the other excluding it, so as to 
appreciate how well the background underneath the SN was fit. 
\textbf{Lower panel:} Wavelength-averaged spatial residuals in units of 
the statistical noise of the input 2D spectrum. For all cases we have 
$\delta F_{\rm tot} \leq 1$, meaning the combined SN+background spatial 
profile is restored to the statistical noise limit over the whole spatial 
range.
              }
   \label{specinhofig1}
\end{figure*}

\begin{figure*}
\centering   	
   \includegraphics[width=18cm]{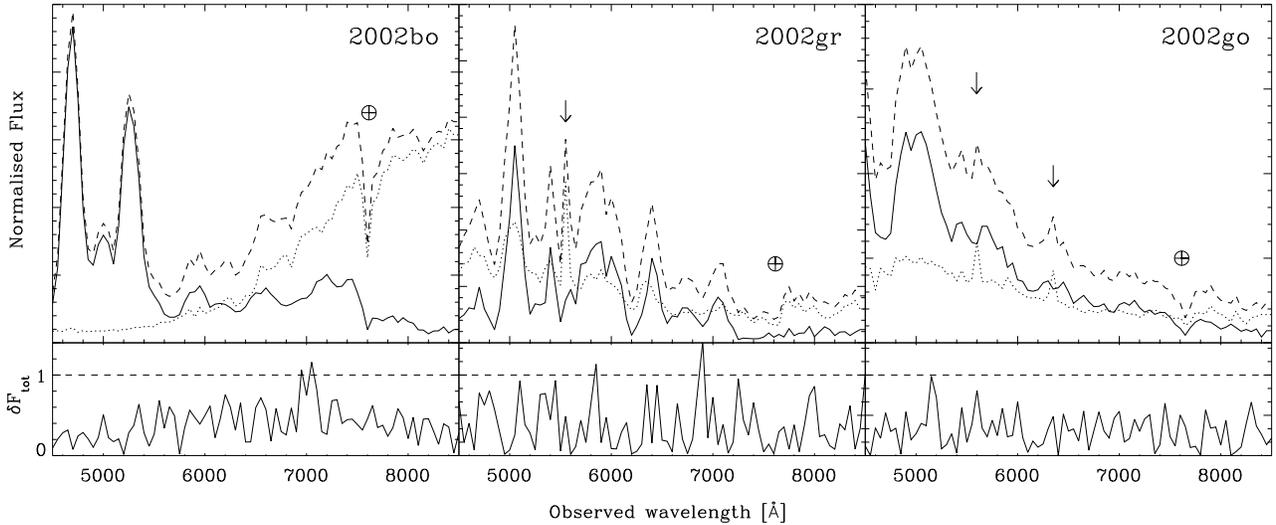}
   \caption{\textbf{Top panel:} Normalised restored point source spectra 
(solid line) and underlying background, both including (dashed line) and 
excluding (dotted line) the point source. The spectra have been normalised to 
the integral of the underlying background flux. The symbol $\oplus$ denotes 
the atmospheric A-band ($\sim$7600-7630\AA), whilst arrows indicate strong 
sky emission lines. \textbf{Lower panel:} Spectral residuals in units of 
the statistical noise of the input 2D spectrum. 
              }
   \label{specinhofig2}
\end{figure*}

In this section we apply this restoration technique to real supernova data. 
An outline of the steps involved (determination of the SN trace position and 
FWHM; SSF synthesis; running of \emph{specinholucy}) has already been given 
in section \ref{simsteps}. We simply provide the input parameters that were 
used to run the code (see Table \ref{specinhoparams}). All the data were 
collected by the authors via the ESO observing programme 170.A-0519 and are 
available through the ESO Science Archive 
Facility\footnote{http://archive.eso.org/}. The observations were made with 
VLT+FORS1 in Long Slit Spectroscopy (LSS) mode with a 300V grism and a 
1\arcsec\ slit. The results of the two-channel restoration are shown in Figs. 
\ref{specinhofig1} and \ref{specinhofig2}, whilst the individual objects are 
presented in Table \ref{sndatasummary}. On average $\sim 150$ iterations were 
needed to reach convergence for the restoration of the point source spectra, 
corresponding to $\sim2$ minutes on a 2.4 GHz Pentium 4 processor.

There is one important distinction to be made in the calculation of the flux 
residuals here as opposed to that made for the simulation in section 
\ref{statcalc}. For observed cases we do not know how the total flux 
$F_{\rm tot}$ separates into the individual components $F_{\rm SN}$, 
$F_{\rm gal}$ and $F_{\rm sky}$, and so our evaluation of the accuracy of the 
restoration is restricted to the calculation of the mean \emph{total} flux 
residuals $\delta F_{\rm tot}$:

\begin{displaymath}
\delta F_{\rm tot} = \frac{ \overline{|F_{\rm tot,i} - F_{\rm tot,r}|}} 
{ \sqrt{ F_{\rm tot,i} }}.
\end{displaymath}	

\noindent
This is not as limiting as it may seem since we are again able to 
quantitatively evaluate our combination of $\sigma_{\rm kernel}$ and 
$\sigma_{\rm SSF}$ through maximising the counts in the point source channel 
whilst ensuring that $\delta F_{\rm tot} < 1$ (cf. section \ref{spatialres}).

Fig. \ref{specinhofig1} illustrates what has been said about an 
\emph{a posteriori} check of the adequacy of both the SSF and the spatial 
resolution kernel. In all cases we not only fit the SN trace but also the 
underlying background such that the flux residuals 
$\delta F_{\rm tot} \leq 1$. For the case of SN 2002bo a secondary point 
source (indicated by an arrow) in its vicinity had to be included in the input
 position table -- and hence allocated to the point source channel -- for the 
restoration to succeed. 

Fig. \ref{specinhofig2} on the other hand shows the flux residuals in the 
dispersion direction, and one can immediately appreciate the successful 
allocation of (extended) atmospheric absorption features and sky emission 
lines to the background channel. Moreover one can immediately pick out the 
spectral regions affected by systematic errors in the restoration 
($\delta F_{\rm tot} > 1$) and only use those where 
$\delta F_{\rm tot} \leq 1$ for analysis. The (spectral) residuals in this 
case were evaluated in the dispersion direction by evaluating the 
\emph{spatial} residuals in a region centred on the SN position and of extent
 $3\sigma_{\rm SSF}$. This ensures that we are only including the point source
 and its immediately underlying background.

\begin{table}
\caption{\emph{specinholucy} parameters}
\label{specinhoparams}
\centering
\begin{tabular}{l c c c l l}
\hline\hline
SN$^{\rm a}$	& $\sigma_{\rm kernel}$	& \multicolumn{2}{c}{FWHM} & SN+bg$^{\rm c}$ & SN/bg$^{\rm d}$ \\ 
		& [pixels]		& DIMM      & SSF          &                 &                 \\     	
\hline
2002bo\_1	& 4.0	& 6.25	& 6.93	& 63.6	& 4.8 	\\ 
2002bo\_2	& 4.0	& 7.95	& 6.57	& 62.5	& 4.6 	\\ 

2002gr		& 2.2	& 5.85	& 3.46	& 9.0	& 1.4 	\\
					   
2002go		& 3.5	& 5.75	& 3.97	& 5.2	& 2.1 	\\
\hline
\end{tabular}
{\footnotesize
\flushleft
(a) IAU designation. Note that two spectra were taken of SN 2002bo.\\
(b) mean FWHM in pixels of the gaussian seeing corresponding to the DIMM station measurement or to the SSF used to extract the spectrum. \\
(c) mean number of counts (ADU) in the dispersion direction of the input 2D spectrum at the location of the SN.\\
(d) ratio of the integrated SN flux and that of the underlying background. One can calculate the mean signal-to-noise per pixel using ${\rm SNR\ pix}^{-1} = {\rm SN/bg}+1$. \\
}
\end{table}

\begin{table}
\caption{Supernova observation summary}
\label{sndatasummary}
\centering
\begin{tabular}{l l l l l}
\hline \hline
SN$^{\rm a}$	& Archive id.$^{\rm b}$& UT date	& $z^{\rm c}$ & phase$^{\rm d}$ \\
\hline
2002bo	 & 2002bo   & 06 Dec 2002 & 0.0042  & +258		\\
2002gr	 & sloan9   & 11 Oct 2002 & 0.091   & +14$^{\dag}$	\\
2002go 	 & sloan5   & 11 Oct 2002 & 0.236   & +2$^{\dag}$	\\	
\hline
\end{tabular}
{\footnotesize
\flushleft
(a) IAU designation. \\
(b) target name for search in the ESO archive. \\
(c) redshift determined from nebular lines in the host galaxy. \\
(d) SN phase in rest-frame days from $B$-band maximum. Phases marked with 
$^{\dag}$ have been determined from cross-correlations with local SN Ia 
templates using the SuperNova IDentification (SNID) code \citep{tonry05}.\\
}
\end{table}

\section{Comparison with other methods \label{comparison}}

\begin{figure*}
\centering   	
   \includegraphics[width=18cm]{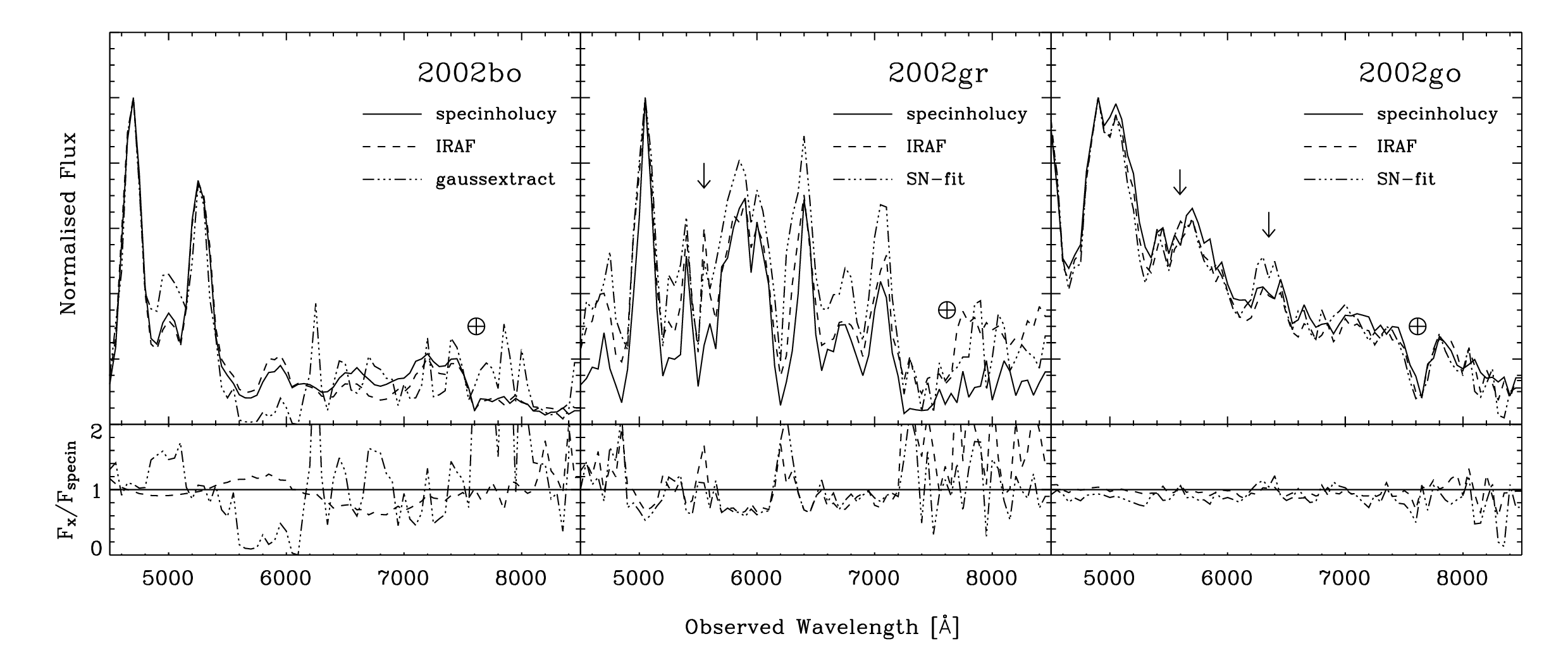}
   \caption{\textbf{Top panel:} Restored supernova spectra using 
\emph{specinholucy} (solid line) and other methods presented in section 
\ref{comparison}. The spectra are normalised to the maximum flux value of the 
\emph{specinholucy} output. Due to the specificities of 
$\mathcal{SN}$-fit -- namely the restriction of the local SN template spectra 
to epochs $-20{\rm d}<t<+20{\rm d}$, we restrict its application to SNe 2002gr
 and 2002go (we plot the solution corresponding to the smallest $\chi^2$ -- 
see text.), for which we have not plotted the \emph{gaussextract} output for
 sake of clarity. The symbol $\oplus$ and arrows have the same meaning as in 
Fig. \ref{specinhofig2}. \textbf{Lower panel:} Ratios of the various 
spectra to the \emph{specinholucy} output. We have deliberately restricted 
the y-axis range to [0,2] to be able to visualise small $\sim 20\%$ 
differences in the SN flux.
              }
   \label{compare}
\end{figure*}

We choose to compare our method with alternative techniques, namely a standard
 extraction using IRAF, an iterative gaussian extractor and a statistical 
decomposition of a 1D flux-calibrated spectrum into SN and galaxy 
contributions. The reason for including the latter (1D) approach is to 
highlight the need for the SN-galaxy separation to occur at the earliest 
stages of the reduction process, and not as part of a post-processing chain of
 data analysis. The results are summarised in Fig. \ref{compare}. Note that 
algorithms similar to ours do exist (though unlike ours are not publicly 
available), and should these be of interest to the reader we refer him/her 
to \citet{LW03}, section 6.

		\subsection{Standard extraction in IRAF}

For most purposes IRAF and similar reduction software are sufficient for 
extracting a high S/N supernova spectral trace. The extraction can further be 
optimised by e.g. tracing the point source signal along the dispersion 
direction or variance-weighting the output flux. However, the hold on 
\emph{systematic} errors is null and one tends to evaluate the resulting 
spectrum on purely qualitative grounds. In the case of high-$z$ SN Ia 
spectroscopy the extraction of the SN spectrum is often considered successful 
when the cross-correlation with a local SN Ia template spectrum is maximal. 
This clearly bypasses the specificity of the high-$z$ spectrum and affects the
 search for systematic spectral differences between high-$z$ SN Ia spectra and
 their local counterparts.

Of equal concern is the arbitrary definition of the SN ``background'' in IRAF.
 One specifies a set of spatial coordinates in the vicinity of the SN trace to
 be fit by some user-defined polynomial \ldots\ and essentially hopes for the 
best! In particular one has to find a compromise between the varying width of 
the SN cross-dispersion profile (i.e. its FWHM-wavelength relation) and the 
varying structure of the underlying background. Cases such as that encountered
 for SN 2002bo -- namely the contamination of the close SN background by a 
secondary point source -- are extremely delicate to handle in a coherent way 
with standard spectral reduction packages and lead in this case to 
$\gtrsim 40\%$ errors in the SN flux. Other errors in the extraction using 
IRAF are due to the varying spectrum of the underlying background, to strong 
sky emission lines and to the overall increase in sky noise at long 
wavelengths. The extracted spectrum of SN 2002go for instance is clearly 
affected by sky residuals, and more specifically the Si II feature (6355\AA) 
blueshifted to $\sim 6150$\AA\ and characteristic of Type Ia supernovae lies in
 the same region as the atmospheric A-band at $z=0.236$, and is clearly 
contaminated by it. Any measure of line strengths and ratios in this region 
will thus also be affected, leading to significant errors in empirical 
correlations based on such measurements, e.g. the ``Nugent relation'' (see 
\citealt{nugent95} and section \ref{impactssf} in this paper). In cases where 
the signal-to-noise ratio is low (cf. SN 2002gr) there is obvious 
contamination from the underlying background in the SN spectrum, leading to 
significant errors; in fact, no SN flux at all is output in many bandpasses, 
where \emph{specinholucy} is able to restore the SN contribution over the 
whole wavelength range.

	\subsection{\emph{Gaussextract} -- an iterative gaussian extractor}

In applying \emph{specinholucy} to the restoration of supernova spectra we 
need to degrade the resolution of the background channel to discriminate 
between the SN and its underlying background. To illustrate the advantages of 
using a two-channel restoration we have written a software 
(\emph{gaussextract}) that iteratively extracts the point source spectrum 
using the \emph{same} synthetic SSF as for the \emph{specinholucy} run. As 
in \emph{specinholucy} each dispersion channel is treated independently and 
the process is iterated till some minimal fractional change in the restored SN
 flux is achieved. For the sake of comparison with \emph{specinholucy}, 
\emph{gaussextract} runs the same convergence criteria for the 
point source flux. The difference here is that the SN and underlying 
background are jointly extracted in the same channel, leading to significant 
contamination by sky lines and host galaxy in the output SN spectrum (cf. 
SN 2002bo in Fig.\ref{compare}).

	\subsection{Statistical approach using $\mathcal{SN}$-fit\label{snfit}}

$\mathcal{SN}$-fit is a software developed by Gr\'egory Sainton 
\citep[see][]{saintonphd} to rapidly identify the supernova
type and redshift in the SuperNova Legacy Survey (SNLS). The goal is to 
quickly identify 
candidates from their spectrum, for which the SN phase
 and the fraction of host galaxy light can be estimated. The algorithm is 
based on a $\chi^2$ fitting of the observed (and reduced) 1D 
spectrum to known template spectra. More specifically the observed
spectrum is compared with a model made of a fraction $\alpha$ of SN
($\mathcal{S}$) and a fraction $\beta$ of galaxy ($\mathcal{G}$). The
following model spectra $\mathcal{M}$ are built :

\begin{eqnarray*}
    \mathcal{M}(\lambda_{\rm rest})(z,\alpha,\beta) = 
\alpha\mathcal{S}(\lambda_{\rm rest}(1+z)) + 
    \left\{
    \begin{array}{ll}	
    \beta\mathcal{G}(\lambda_{\rm rest}(1+z))& \textrm{(a)}\\
    \beta\mathcal{G}(\lambda_{\rm obs}) & \textrm{(b),}    
    \end{array}	
    \right.	
\end{eqnarray*}

\noindent
where $\lambda_{\rm rest}$ and $\lambda_{\rm obs}$ are the rest-frame and 
observed wavelength, respectively. (a) corresponds to the case where a local 
galaxy template spectrum is redshifted to fit the observed spectrum, whereas 
(b) makes use of the observed SN host galaxy itself. 
The $\chi^2$ is made robust against aberrant
points (remaining sky lines, mostly). The fit is done for every 
galaxy-supernova pair and the results are sorted in ascending $\chi^2$. One is
 therefore able to evaluate how 
significant is the best result with respect to other SN-galaxy combinations, 
and in turn how confident one should be about the output supernova type, 
redshift and epoch. 

It should be noted that, contrary to the other methods presented in this 
section, $\mathcal{SN}$-fit is a \emph{post}-processing tool and thus cannot 
be required to correct for systematic errors made in the reduction process 
(most likely using standard software packages such as IRAF or 
MIDAS\footnote{ESO-MIDAS is a registered trademark of the European Southern 
Observatory; see http://www.eso.org/projects/esomidas/.}), such as improper 
removal of sky lines. These errors will propagate in the $\mathcal{SN}$-fit 
result, as seen in the output spectrum of SN 2002go where the Si II feature is
 also affected by the atmospheric A-band. SN 2002gr on the other hand 
illustrates the limits of using a $\chi^2$ minimisation technique to separate 
the SN from the underlying background in low signal-to-noise cases. 

The advantages of this purely statistical approach is that no 
\emph{a priori} assumption is made about the underlying background. The host
 galaxy contamination is removed using a galaxy template or better still the 
SN host galaxy itself. The determination of the supernova type and redshift is
 greatly improved upon and the enhanced accuracy of this approach is discussed
 in \citet{saintonphd}. The disadvantage is that we are limited by the 
database of local SNe, and any spectral peculiarities will not be 
independently accounted for, if they are not reproduced in local SN spectra. 
Also, the fact that in both cases we use the observed SN host galaxy itself 
as a template to decompose the extracted SN trace into a supernova and a 
galaxy component -- case (b) in the above system of equations -- illustrates 
what we were implying in section \ref{spatialres} about the propagation of 
supernova ``features'' in the final SN spectrum due to the inherent 
imperfections of the galaxy template removal approach.


\section{Conclusion \label{conclusion}}

We have presented the advantages of using a more elaborate spectral extraction
 algorithm for obtaining supernova spectra devoid of background contamination,
 and hence enable a truely quantitative analysis of high-$z$ Type Ia supernova
 spectra. This will allow us in the future to set constraints on potential 
evolutionary effects that could affect SNe Ia as function of redshift. Not 
only are we able to extract the PSF-like components of an input spectrum, one 
can also restore the entire 2D spectrum and hence get a hold on systematic 
errors linked with the restoration process. We thus have a quantitative way of
 judging whether or not the extracted SN spectrum is optimal for subsequent 
analysis and whether we are not mistakenly ``seeing'' SN-like features in a 
pure galaxy spectrum, as is sometimes the case in low S/N high-$z$ SN Ia 
spectra.

However, there are some caveats mainly associated with the complexity of 
\emph{specinholucy}. Errors associated with the use of an inappropriate SSF 
for restoring the point-source channel or a mis-evaluation of the width of the
 spatial resolution kernel necessary for the point-source/extended background 
discrimination will strongly affect the restoration (see section 
\ref{ssfissue}). Through restoration of the 2D background spectrum one is 
however able to produce diagnostic plots to critically evaluate the error in 
each case and repeat the restoration with different settings (see Fig. 
\ref{skernel}). 

Note that a valuable by-product of this novel technique is the acquisition of 
a host galaxy spectrum with no contamination from the supernova flux! Galaxies
 at $z \sim 0.5$ are at most a few arcseconds in size and one can essentially 
obtain integrated spectra of high-$z$ SN host galaxies by centering a slit on 
both the SN and its host. This in turn enables the study of properties of 
high-$z$ SN Ia host galaxies and compare them with integrated spectra of local
 SN Ia hosts \citep{sullivan03,williams03,gallagher04}.

Thus, \emph{specinholucy} provides the high-$z$ supernova community with a 
reliable tool to obtain clean SN Ia spectra and in turn to derive accurate 
physical quantities associated with the ejecta. Two of the authors 
(St\'ephane Blondin and Bruno Leibundgut) are part of the ESSENCE project, 
and Gr\'egory Sainton is part of the SNLS collaboration. Both teams recognise 
the need for a quantitative analysis of their results in order to constrain 
potential systematic effects that could distort the cosmological results 
derived from observations of high-$z$ SNe Ia. We have shown 
\emph{specinholucy} to be the ideal tool for this purpose. In fact, the VLT 
data taken for the ESSENCE project will be shown in their restored form via 
\emph{specinholucy} in \citealt{matheson05}.


\begin{acknowledgements}
Many thanks to John Moustakas for providing the zero-point for the spectrum of
 NGC 6181 and to Mattia Vaccari for deriving the galaxy luminosity profile 
used in the simulation. We further would like to thank the anonymous referee 
for many detailed and useful comments on the original manuscript.
\end{acknowledgements}

\bibliographystyle{aa}
\bibliography{specinholucy}

\begin{thebibliography}{41}
\expandafter\ifx\csname natexlab\endcsname\relax\def\natexlab#1{#1}\fi

\bibitem[{{Avila} {et~al.}(1997){Avila}, {Rupprecht}, \& {Beckers}}]{avila97}
{Avila}, G., {Rupprecht}, G., \& {Beckers}, J.~M. 1997, in Proc. SPIE Vol.
  2871, p. 1135-1143, Optical Telescopes of Today and Tomorrow, Arne L.
  Ardeberg; Ed., 1135--1143

\bibitem[{{Barris} {et~al.}(2004){Barris}, {Tonry}, {Blondin}, {Challis},
  {Chornock}, {Clocchiatti}, {Filippenko}, {Garnavich}, {Holland}, {Jha},
  {Kirshner}, {Krisciunas}, {Leibundgut}, {Li}, {Matheson}, {Miknaitis},
  {Riess}, {Schmidt}, {Smith}, {Sollerman}, {Spyromilio}, {Stubbs}, {Suntzeff},
  {Aussel}, {Chambers}, {Connelley}, {Donovan}, {Henry}, {Kaiser}, {Liu},
  {Mart{\'{\i}}n}, \& {Wainscoat}}]{barris04}
{Barris}, B.~J., {Tonry}, J.~L., {Blondin}, S., {et~al.} 2004, \apj, 602, 571

\bibitem[{{de Vaucouleurs}(1959)}]{deVaucouleurs}
{de Vaucouleurs}, G. 1959, Handbuch der Physik, 53, 275

\bibitem[{{Drenkhahn} \& {Richtler}(1999)}]{ngc4526}
{Drenkhahn}, G. \& {Richtler}, T. 1999, \aap, 349, 877

\bibitem[{{Filippenko}(1997)}]{filippenko97}
{Filippenko}, A.~V. 1997, \araa, 35, 309

\bibitem[{{Freeman}(1970)}]{freeman70}
{Freeman}, K.~C. 1970, \apj, 160, 811

\bibitem[{{Fried}(1965)}]{fried65}
{Fried}, D.~L. 1965, Optical Society of America Journal, 55, 1427

\bibitem[{{Fried}(1975)}]{fried75}
{Fried}, D.~L. 1975, Radio Science, 10, 71

\bibitem[{{Ghallager} \& {Garnavich}(2004)}]{gallagher04}
{Ghallager}, S. \& {Garnavich}, P. 2004, in prep

\bibitem[{{Hillebrandt} \& {Niemeyer}(2000)}]{wolfi00}
{Hillebrandt}, W. \& {Niemeyer}, J.~C. 2000, \araa, 38, 191

\bibitem[{{Hook} \& {Lucy}(1994)}]{hooklucy94}
{Hook}, R.~N. \& {Lucy}, L.~B. 1994, in The Restoration of HST Images and
  Spectra - II, 86--+

\bibitem[{{Kennicutt}(1992)}]{kenn1}
{Kennicutt}, R.~C. 1992, \apjs, 79, 255

\bibitem[{{Kent}(1985)}]{kent85}
{Kent}, S.~M. 1985, \apjs, 59, 115

\bibitem[{{Knop} {et~al.}(2003){Knop}, {Aldering}, {Amanullah}, {Astier},
  {Blanc}, {Burns}, {Conley}, {Deustua}, {Doi}, {Ellis}, {Fabbro}, {Folatelli},
  {Fruchter}, {Garavini}, {Garmond}, {Garton}, {Gibbons}, {Goldhaber},
  {Goobar}, {Groom}, {Hardin}, {Hook}, {Howell}, {Kim}, {Lee}, {Lidman},
  {Mendez}, {Nobili}, {Nugent}, {Pain}, {Panagia}, {Pennypacker}, {Perlmutter},
  {Quimby}, {Raux}, {Regnault}, {Ruiz-Lapuente}, {Sainton}, {Schaefer},
  {Schahmaneche}, {Smith}, {Spadafora}, {Stanishev}, {Sullivan}, {Walton},
  {Wang}, {Wood-Vasey}, \& {Yasuda}}]{knop03}
{Knop}, R.~A., {Aldering}, G., {Amanullah}, R., {et~al.} 2003, \apj, 598, 102

\bibitem[{{Leibundgut}(2001)}]{bruno01}
{Leibundgut}, B. 2001, \araa, 39, 67

\bibitem[{{Lucy}(1974)}]{lucy74}
{Lucy}, L.~B. 1974, \aj, 79, 745

\bibitem[{{Lucy}(1994)}]{lucy94}
{Lucy}, L.~B. 1994, in The Restoration of HST Images and Spectra - II, 79--+

\bibitem[{{Lucy} \& {Walsh}(2003)}]{LW03}
{Lucy}, L.~B. \& {Walsh}, J.~R. 2003, \aj, 125, 2266

\bibitem[{{Matheson} {et~al.}(2005){Matheson}, {Blondin}, {Fowley}, {Chornock},
  {Filippenko}, {Leibundgut}, {Sollerman}, {Spyromillio}, \& {the ESSENCE
  Team}}]{matheson05}
{Matheson}, T., {Blondin}, S., {Fowley}, R., {et~al.} 2005, \apj, submitted

\bibitem[{{Miknaitis} {et~al.}(2005){Miknaitis}, {Tonry}, {Garnavich},
  {Stubbs}, {Smith}, \& {the ESSENCE Team}}]{miknaitis05}
{Miknaitis}, G., {Tonry}, J., {Garnavich}, P., {et~al.} 2005, \apj, submitted

\bibitem[{{Nugent} {et~al.}(1995){Nugent}, {Phillips}, {Baron}, {Branch}, \&
  {Hauschildt}}]{nugent95}
{Nugent}, P., {Phillips}, M., {Baron}, E., {Branch}, D., \& {Hauschildt}, P.
  1995, \apjl, 455, L147+

\bibitem[{{Pain} \& {SNLS Collaboration}(2002)}]{pain02}
{Pain}, R. \& {SNLS Collaboration}. 2002, Bulletin of the American Astronomical
  Society, 34, 1169

\bibitem[{{Patat}(2003)}]{patat03}
{Patat}, F. 2003, \aap, 400, 1183

\bibitem[{{Patat} {et~al.}(1996){Patat}, {Benetti}, {Cappellaro}, {Danziger},
  {della Valle}, {Mazzali}, \& {Turatto}}]{patat94D}
{Patat}, F., {Benetti}, S., {Cappellaro}, E., {et~al.} 1996, \mnras, 278, 111

\bibitem[{{Perlmutter} {et~al.}(1999){Perlmutter}, {Aldering}, {Goldhaber},
  {Knop}, {Nugent}, {Castro}, {Deustua}, {Fabbro}, {Goobar}, {Groom}, {Hook},
  {Kim}, {Kim}, {Lee}, {Nunes}, {Pain}, {Pennypacker}, {Quimby}, {Lidman},
  {Ellis}, {Irwin}, {McMahon}, {Ruiz-Lapuente}, {Walton}, {Schaefer}, {Boyle},
  {Filippenko}, {Matheson}, {Fruchter}, {Panagia}, {Newberg}, {Couch}, \& {The
  Supernova Cosmology Project}}]{P99}
{Perlmutter}, S., {Aldering}, G., {Goldhaber}, G., {et~al.} 1999, \apj, 517,
  565

\bibitem[{{Phillips}(1993)}]{phillips93}
{Phillips}, M.~M. 1993, \apjl, 413, L105

\bibitem[{{Ratnatunga} {et~al.}(1999){Ratnatunga}, {Griffiths}, \&
  {Ostrander}}]{Ratnatunga}
{Ratnatunga}, K.~U., {Griffiths}, R.~E., \& {Ostrander}, E.~J. 1999, \aj, 118,
  86

\bibitem[{{Richardson}(1972)}]{richardson72}
{Richardson}, W.~H. 1972, Optical Society of America Journal, 62, 55

\bibitem[{{Riess} {et~al.}(1998{\natexlab{a}}){Riess}, {Filippenko}, {Challis},
  {Clocchiatti}, {Diercks}, {Garnavich}, {Gilliland}, {Hogan}, {Jha},
  {Kirshner}, {Leibundgut}, {Phillips}, {Reiss}, {Schmidt}, {Schommer},
  {Smith}, {Spyromilio}, {Stubbs}, {Suntzeff}, \& {Tonry}}]{R98}
{Riess}, A.~G., {Filippenko}, A.~V., {Challis}, P., {et~al.}
  1998{\natexlab{a}}, \aj, 116, 1009

\bibitem[{{Riess} {et~al.}(1998{\natexlab{b}}){Riess}, {Nugent}, {Filippenko},
  {Kirshner}, \& {Perlmutter}}]{riesssnapshot}
{Riess}, A.~G., {Nugent}, P., {Filippenko}, A.~V., {Kirshner}, R.~P., \&
  {Perlmutter}, S. 1998{\natexlab{b}}, \apj, 504, 935

\bibitem[{{Riess} {et~al.}(2004){Riess}, {Strolger}, {Tonry}, {Tsvetanov},
  {Casertano}, {Ferguson}, {Mobasher}, {Challis}, {Panagia}, {Filippenko},
  {Li}, {Chornock}, {Kirshner}, {Leibundgut}, {Dickinson}, {Koekemoer},
  {Grogin}, \& {Giavalisco}}]{riesshz1}
{Riess}, A.~G., {Strolger}, L., {Tonry}, J., {et~al.} 2004, \apjl, 600, L163

\bibitem[{{Sainton}(2004)}]{saintonphd}
{Sainton}, G. 2004, PhD thesis, Universit\'e Lyon I, N$^\circ$ ordre: 131-2004

\bibitem[{{Sandrock} {et~al.}(2000){Sandrock}, {Amestica}, {Duhoux},
  {Navarrete}, \& {Sarazin}}]{sandrock00}
{Sandrock}, S., {Amestica}, R., {Duhoux}, P., {Navarrete}, J., \& {Sarazin},
  M.~S. 2000, in Proc. SPIE Vol. 4009, p. 338-349, Advanced Telescope and
  Instrumentation Control Software, Hilton Lewis; Ed., 338--349

\bibitem[{{Sarazin} \& {Roddier}(1990)}]{sarazin90}
{Sarazin}, M. \& {Roddier}, F. 1990, \aap, 227, 294

\bibitem[{{Schmidt} {et~al.}(1998){Schmidt}, {Suntzeff}, {Phillips},
  {Schommer}, {Clocchiatti}, {Kirshner}, {Garnavich}, {Challis}, {Leibundgut},
  {Spyromilio}, {Riess}, {Filippenko}, {Hamuy}, {Smith}, {Hogan}, {Stubbs},
  {Diercks}, {Reiss}, {Gilliland}, {Tonry}, {Maza}, {Dressler}, {Walsh}, \&
  {Ciardullo}}]{schmidt98}
{Schmidt}, B.~P., {Suntzeff}, N.~B., {Phillips}, M.~M., {et~al.} 1998, \apj,
  507, 46

\bibitem[{{Schroeder}(1987)}]{schroeder}
{Schroeder}, D.~J. 1987, {Astronomical Optics} (San Diego: Academic Press,
  1987)

\bibitem[{{Sil'chenko} {et~al.}(1997){Sil'chenko}, {Zasov}, {Burenkov}, \&
  {Boulesteix}}]{silchenko97}
{Sil'chenko}, O.~K., {Zasov}, A.~V., {Burenkov}, A.~N., \& {Boulesteix}, J.
  1997, \aaps, 121, 1

\bibitem[{{Sullivan} {et~al.}(2003){Sullivan}, {Ellis}, {Aldering},
  {Amanullah}, {Astier}, {Blanc}, {Burns}, {Conley}, {Deustua}, {Doi},
  {Fabbro}, {Folatelli}, {Fruchter}, {Garavini}, {Gibbons}, {Goldhaber},
  {Goobar}, {Groom}, {Hardin}, {Hook}, {Howell}, {Irwin}, {Kim}, {Knop},
  {Lidman}, {McMahon}, {Mendez}, {Nobili}, {Nugent}, {Pain}, {Panagia},
  {Pennypacker}, {Perlmutter}, {Quimby}, {Raux}, {Regnault}, {Ruiz-Lapuente},
  {Schaefer}, {Schahmaneche}, {Spadafora}, {Walton}, {Wang}, {Wood-Vasey}, \&
  {Yasuda}}]{sullivan03}
{Sullivan}, M., {Ellis}, R.~S., {Aldering}, G., {et~al.} 2003, \mnras, 340,
  1057

\bibitem[{{Tonry} {et~al.}(2005){Tonry}, {Salvo}, {Blondin}, {Schmidt}, \&
  {Leibundgut}}]{tonry05}
{Tonry}, J.~L., {Salvo}, M., {Blondin}, S., {Schmidt}, B.~P., \& {Leibundgut},
  B. 2005, in prep

\bibitem[{{Tonry} {et~al.}(2003){Tonry}, {Schmidt}, {Barris}, {Candia},
  {Challis}, {Clocchiatti}, {Coil}, {Filippenko}, {Garnavich}, {Hogan},
  {Holland}, {Jha}, {Kirshner}, {Krisciunas}, {Leibundgut}, {Li}, {Matheson},
  {Phillips}, {Riess}, {Schommer}, {Smith}, {Sollerman}, {Spyromilio},
  {Stubbs}, \& {Suntzeff}}]{tonry03}
{Tonry}, J.~L., {Schmidt}, B.~P., {Barris}, B., {et~al.} 2003, \apj, 594, 1

\bibitem[{{Williams} {et~al.}(2003){Williams}, {Hogan}, {Barris}, {Candia},
  {Challis}, {Clocchiatti}, {Coil}, {Filippenko}, {Garnavich}, {Kirshner},
  {Holland}, {Jha}, {Krisciunas}, {Leibundgut}, {Li}, {Matheson}, {Maza},
  {Phillips}, {Riess}, {Schmidt}, {Schommer}, {Smith}, {Sollerman},
  {Spyromilio}, {Stubbs}, {Suntzeff}, \& {Tonry}}]{williams03}
{Williams}, B.~F., {Hogan}, C.~J., {Barris}, B., {et~al.} 2003, \aj, 126, 2608

\end{thebibliography}

\end{document}